\documentclass[twocolumn]{aastex631}
\usepackage{apjfonts}
\usepackage{rotating}
\usepackage{amsmath, color,graphicx,tablefootnote,subfigure}
\usepackage{graphics,subfigure,hyperref,float}
\usepackage[rightcaption]{sidecap}

\newcommand{\CH}[1]{\colhead{#1}}
\newcommand{\D}{$^{\dagger}$}

\newcommand{\CM}{\checkmark}
\newcommand{\W}{$\lambda$}

\shorttitle{CHAOS VIII}
\shortauthors{Berg et al.}

\begin{document}

\shortauthors{Berg et al.}
\title{CHAOS {\sc viii}: Far-Ultraviolet Spectra of M101 and The Impact of Wolf-Rayet Stars\footnote{
Based on observations made with the NASA/ESA Hubble Space Telescope,
obtained from the Data Archive at the Space Telescope Science Institute, which
is operated by the Association of Universities for Research in Astronomy, Inc.,
under NASA contract NAS 5-26555.}}

\author[0000-0002-4153-053X]{Danielle A. Berg}
\affiliation{Department of Astronomy, The University of Texas at Austin, 2515 Speedway, Stop C1400, Austin, TX 78712, USA}
\author[0000-0003-0605-8732]{Evan D.\ Skillman}
\affiliation{Minnesota Institute for Astrophysics, University of Minnesota, 116 Church St. SE, Minneapolis, MN 55455}
\author[0000-0002-0302-2577]{John Chisholm}
\affiliation{Department of Astronomy, The University of Texas at Austin, 2515 Speedway, Stop C1400, Austin, TX 78712, USA}
\author[0000-0003-1435-3053]{Richard W.\ Pogge}
\affiliation{Department of Astronomy, The Ohio State University, 140 W 18th Ave., Columbus, OH, 43210}
\affiliation{Center for Cosmology \& AstroParticle Physics, The Ohio State University, 191 West Woodruff Avenue, Columbus, OH 43210}
\author[0000-0002-5659-4974]{Simon Gazagnes}
\affiliation{Department of Astronomy, The University of Texas at Austin, 2515 Speedway, Stop C1400, Austin, TX 78712, USA}
\author[0000-0002-0361-8223]{Noah S.\ J.\ Rogers}
\affiliation{Department of Physics and Astronomy, Northwestern University, 2145 Sheridan Road, Evanston, IL 60208, USA}
\affiliation{Center for Interdisciplinary Exploration and Research in Astrophysics (CIERA), Northwestern University, 1800 Sherman Avenue, Evanston, IL 60201, USA}
\author[0000-0001-9714-2758]{Dawn K.\ Erb}
\affiliation{Center for Gravitation, Cosmology and Astrophysics, Department of Physics, University of Wisconsin Milwaukee, 3135 N Maryland Ave., Milwaukee, WI 53211, USA}
\author[0000-0002-2644-3518]{Karla Z. Arellano-C\'{o}rdova}
\affiliation{Institute for Astronomy, University of Edinburgh, Royal Observatory, Edinburgh, EH9 3HJ, UK}
\author[0000-0003-2685-4488]{Claus Leitherer}
\affiliation{Space Telescope Science Institute, 3700 San Martin Drive, Baltimore, MD 21218, USA}
\author{Jackie Appel}
\affiliation{Associate News Editor, Popular Mechanics Magazine}
\author[0000-0002-2733-4559]{John Moustakas}
\affiliation{Department of Physics \& Astronomy, Siena College, 515 Loudon Road, Loudonville, NY 12211}


\begin{abstract}
We investigate the stellar and nebular properties 
of 9 \ion{H}{2} regions in the spiral galaxy M101 with far-ultraviolet 
(FUV; $\sim900-2000$~\AA) and optical ($\sim 3200-10000$~\AA) spectra.
We detect significant \ion{C}{3}] \W\W1907,1909 nebular emission in 7 regions, but 
\ion{O}{3}] \W1666 only in the lowest-metallicity region.
We produce new analytic functions of the carbon ICF as a function of
metallicity in order to perform a preliminary C/O abundance analysis.
The FUV spectra also contain numerous stellar emission and P-Cygni features that
we fit with luminosity-weighted combinations of single-burst \texttt{Starburst99} and \texttt{BPASS} models.
We find that the best-fit \texttt{Starburst99} models closely match the observed very-high-ionization
P-Cygni features, requiring very-hot, young ($\lesssim3$~Myr),
metal-enriched massive stars.
The youngest stellar populations are strongly correlated with 
broad \ion{He}{2} emission, nitrogen Wolf-Rayet (WR) FUV and optical spectral features, and
enhanced N/O gas abundances.
Thus, the short-lived WR phase may be driving excess emission in several N P-Cygni wind 
features (\W955, \W991, \W1720) that bias the stellar continuum fits to higher metallicities 
relative to the gas-phase metallicities.
Accurate characterization of these \ion{H}{2} regions requires additional inclusion of WR 
stars in the stellar population synthesis models.
Our FUV spectra demonstrate that the $\sim900-1200$ \AA\ FUV can provide a strong
test-bed for future WR atmosphere and evolution models.
\end{abstract}

\keywords{Chemical abundances (224), Wolf-Rayet stars (1806), Ultraviolet spectroscopy (2284), H II regions (694), Spiral galaxies (1560)}


\section{Introduction}\label{sec1}
The history of a galaxy can be traced from its cumulative abundance pattern,
as abundances of heavy elements increase with time as successive generations of stars 
return their newly synthesized elements to the interstellar medium (ISM). 
Star-forming regions (\ion{H}{2} regions) are privileged sites to measure 
the chemical abundances in galaxies, near and far, due to their ubiquitous 
strong emission lines in the rest-frame optical regime. 
While the majority of nebular studies come from optical observations,
the far-ultraviolet (FUV; defined here as 1200 -- 2000 \AA) is a rich wavelength
regime with many nebular diagnostic features that are complementary to
those in the optical.  
Recently, the {\it James Webb Space Telescope} ({\it JWST}) has enabled rest-frame 
optical and ultraviolet (UV) spectra of intermediate and high redshift galaxies, 
ushering in a new era of chemical abundance studies across cosmic redshifts
\citep[e.g.,][]{arellano-cordova22,curti23,maseda23,nakajima23,rogers23}.
Robust abundance measurements are key to deciphering 
the complex physical and evolutionary processes of galaxy formation and growth. 
In this regard, securing details of present-day abundances in star forming regions, 
which better allow us to interpret the higher redshift spectra, is crucial for forming
the basis of our understanding of the chemical evolution of the universe. 

Different elements trace different portions of chemical evolution owing to the 
different times scales of their nucleosynthetic production.  
On short time scales ($< 10$ Myr), massive stars ($>10\ M_\odot$) forge heavier elements 
through the $\alpha$-process which fuses $\alpha$-particles into successfully more massive 
elements to create oxygen, neon, magnesium, silicon, and beyond up to iron. 
When these massive stars reach the iron-peak they explode as supernova and deposit 
these $\alpha$-enriched elements into the ISM. 
Carbon and nitrogen, crucial elements for the formation of dust and regulation 
of ISM conditions, are not significantly produced by the $\alpha$-process. 
Rather, the primary mechanism to produce nitrogen and carbon is thought to be neutron 
capture within asymptotic giant branch (AGB) stars, occuring on longer timescales. 
AGB stars are evolved lower mass (0.5--8~M$_\odot$) stars with inert cores of carbon 
and oxygen. 
These AGB stars are thought to be the primary source of nitrogen at low-metallicity 
(12+log(O/H) $\lesssim$ 8.0; $Z\lesssim0.2\ Z_\odot$), with enrichment timescales 
\citep[$\sim250$ Myr;][]{henry00} that occur significantly after the most massive stars have exploded as 
supernovae. 

At higher metallicity ($Z\gtrsim0.2\ Z_\odot$), the carbon-nitrogen-oxygen (CNO) 
cycle provides a ``secondary" source of N, meaning that it is metallicity-dependent.
The CNO cycle uses trace amounts of carbon and oxygen as catalysts in a chain of reactions 
that convert H to He.
Nitrogen is mostly produced in CN branch of the CNO cycle, which uses $^{12}$C as a 
reaction catalyst and creates $^{14}$N as a byproduct.
The slowest ``bottleneck" step of this branch is the $^{14}$N($p, \gamma$)O$^{15}$ step
because it has a relatively low cross-section, allowing   
nitrogen to build up in the cores of high-metallicity massive stars. 
Convection can then dredge up nitrogen to the surface, where stellar winds can advect it
into the ISM.
This process can enhance N-production during certain phases of massive star
evolution, such as the Wolf Rayet (WR) star phase, where stellar winds are 
especially strong. 
Measuring the relative chemical abundance (e.g. N/O or C/O) describes which of these processes a galaxy has undergoing and what the past star formation history of the galaxy looks like. 

Recent observations of high-redshift galaxies have raised questions about the
standard view of chemical abundances. 
For instance, deep rest-frame FUV observations of the galaxy GN-z11 find strong 
\ion{N}{4}] \W1486, \ion{N}{3}] \W1750, and \ion{C}{3}] \W\W1907,1909 emission lines that 
suggest that the galaxy has been enriched to super solar nitrogen abundances just a few 
100~Myr after the Big Bang \citep[e.g.,][]{maiolino23,charbonnel23,senchyna24}.
This is far too rapid for enrichment by AGB stars. 
Previous studies suggest that either WR, super massive stars, tidal disruption events,
or active galactic nuclei environments could be responsible for the high nitrogen abundances
\citep[e.g.,][]{matsuoka17,cameron23,kobayashi24}.
However, the nitrogen and carbon lines observed within GN-z11 are rarely detected in local 
star-forming galaxies, such that a complete understanding of the production and origin of 
these features has not fully been revealed. 
The origin of these features requires a multi-wavelength approach that incorporates 
both the stellar and nebular physics. 

UV spectra offer a unique glimpse into the processes occurring 
in star-forming galactic environments. 
In particular, the FUV regime 
contains a rich diversity of features that can simultaneously
characterize the ionizing sources, feedback energetics, and 
gas conditions within galaxies.
FUV stellar spectra of ionized regions (\ion{H}{2}) trace the massive star population, as 
they are dominated by P-Cygni stellar wind features, composed of a combination of broad, 
blueshifted absorption and redshifted emission.
The so-called stellar wind P-Cygni profiles are produced by strong, 
radiatively-driven stellar photosphere winds \citep[e.g.,][]{castor75, lamers99}. 
Winds from O-stars are usually ionized to the N$^{+3}$, C$^{+4}$, and Si$^{+4}$
ionization states \citep{lamers99}. 
However, none of these ionization states have strong resonant lines blueward of 
Ly$\alpha$ (note that \ion{N}{4} \W 1720 is a non-resonant line).
The lines that are available, \ion{N}{5} \W\W1238,1246, \ion{Si}{4} \W\W1393,1403, 
and \ion{C}{4} \W\W1548,1550, trace adjacent ionization stages. 
As such, these stellar wind signatures are highly sensitive to the temperature 
(and, therefore, the age), density, and metallicity of the winds of the most 
massive / luminous stars. 
This sensitive dependency on stellar properties has important ramifications for 
which lines trace which stars. 
For instance, the \ion{Si}{4} \W\W1392,1402 doublet traces lower ionized gas 
than the peak ionization, such that \ion{Si}{4} is typically only observed in 
populations with cooler stars, such as evolved Giants or B-stars. 
The \ion{N}{5} \W\W1238,1242 doublet on the other hand traces more highly ionized 
gas and is only found in the hottest / youngest stellar populations. 

In this paper we present HST/COS FUV spectral observations for 9 
\ion{H}{2} regions in M101 in \S~\ref{sec:sample} and use these spectra to explore 
the stellar features and nebular emission.
M101, also known as NGC~5457, is an invaluable target for investigating the stellar and
nebular properties across spiral galaxies. 
Its utility stems from its proximity, allowing for detailed, spatially-resolved 
investigations, and its face-on orientation, which presents an unobstructed view of its 
spiral arms and associated nebulae, providing a comprehensive look at the distribution 
and composition of stellar and nebular properties. 
Our sample of \ion{H}{2} regions and spectral observations are described in 
Sections~\ref{sec:sample} and \ref{sec:obs}, respectively. 
In Section~\ref{sec:contfit}, we perform stellar continuum fits to determine the 
average properties of the massive star populations. 
We discuss the FUV spectral characteristics of this sample in Section~\ref{sec:spectra},
including the stellar continuum properties in \S~\ref{sec:continuum} and
the presence of Wolf-Rayet features and \ion{He}{2} emission in \S~\ref{sec:HeII},
In \S~\ref{sec:UV_CO} we examine the nebular C and O emission features and C ionization 
correction factors (ICFs) from our grid of photoionization models in order to perform 
an exploratory FUV C/O abundance analysis and compare to optical C/O measurements.
Our conclusions are given in Section~\ref{sec:conclusions}.


\section{The M101 \ion{H}{2} Region Sample\label{sec:sample}}

For the present study, we targeted the star-forming regions identified by 
\citet{skillman20} as very high surface brightness \ion{H}{2} regions in M101.
In particular, these \ion{H}{2} regions contain detections of $4-5$ intrinsically-faint 
auroral collisionally-excited lines (CELs; see Table~\ref{tbl1}) and the even fainter
\ion{C}{2} \W4267 recombination line (RL) in their
high-quality optical spectra from the CHemical Abundances Of Spirals 
\citep[CHAOS;][]{berg15} project study of M101 \citep{croxall16}.
In total, \citet{skillman20} presented 10 \ion{H}{2} regions. 
For this work, we targeted the same sample of 10 \ion{H}{2} regions,
but one of the UV spectral observations failed, resulting in a sample of
9 \ion{H}{2} regions. 

We note that \citet{skillman20} assumed the same properties for M\,101 as in 
\citet{croxall16}, i.e., a distance of 7.4 Mpc \citep{ferrarese2000}, resulting in 
a spatial scale of 35.9 pc/arcsec; however, \citet{beaton19} have made a (hopefully) 
definitive determination of the distance to M101 of 6.52 $\pm$ 0.12$_{stat}$ $\pm$ 0.15$_{sys}$ Mpc, 
with a resulting scale of 31.6 pc/arcsec. 
We have adopted the updated \citet{beaton19} distance here,
although updates to the distance to M101 will not affect the major conclusions
of this paper.  
We have also assumed disk scalings of 
R$_{25}$ $=$ 864\arcsec\ \citep[radius of $B_{25}$ mag arcsec$^{-2}$;][]{kennicutt11} and 
R$_e = 198$\arcsec\ \citep[half-light radius;][]{berg20},
an inclination angle of 18\degr, and a major-axis position angle of 39\degr\ \citep{walter08}. 

CHAOS uses the Multi-Object Double Spectrographs \citep[MODS,][]{pogge10} on the 
Large Binocular Telescope \citep[LBT,][]{hill10} to observe moderate-resolution 
($R\sim2000$) optical spectra (3,200 \AA\ $\lesssim \lambda \lesssim 10,000$ \AA) of 
a large numbers of \ion{H}{2} regions in spiral galaxies. 
As part of the CHAOS program, very high quality spectra of 74 \ion{H}{2} regions in M101 
\citep{croxall16} were obtained that allowed direct determinations of absolute and relative 
abundances across a broad range of parameter space.
As an extension of the CHAOS project, \citet{skillman20} examined the M101 spectra for 
the optical RL and found detections of the \ion{C}{2} \W4267 line in 10 \ion{H}{2} regions.


\begin{deluxetable*}{cccccc|c|c|c|c|c|c|c}
\tabletypesize{\scriptsize}
\tablecaption{Properties of M101 \ion{H}{2} Regions from the CHAOS Surveys}
\tablehead{
\multicolumn{5}{c}{} & \multicolumn{5}{c}{Auroral Line Detections} & \CH{} & \CH{}\\ \cline{6-10}
\CH{Region}  & \CH{Name}    & \CH{R.A., Decl.} & \CH{Location ID}  & \CH{$\frac{R}{R_{e}}$}  & \CH{12+log(O/H)} &
\CH{[\ion{O}{3}]} & \CH{[\ion{N}{2}]} & \CH{[\ion{S}{3}]} & \CH{[\ion{O}{2}]} & \CH{[\ion{S}{2}]} & \CH{\ion{He}{2}} & \CH{WR}  }
\startdata	
1  & H1013       & 14:03:31.3,+54:21:05.81 & +164.6+009.9      & 0.89     & 8.60$\pm$0.03 & \CM & \CM & \CM & \CM & \CM &     & \CM \\  
2  & H1052       & 14:03:34.1,+54:18:39.60 & +189.2--136.3     & 1.30     & 8.58$\pm$0.02 & \CM & \CM & \CM & \CM & \CM &     & \CM \\  
4  & NGC~5462    & 14:03:53.0,+54:22:06.80 & +354.1+071.2      & 1.89     & 8.51$\pm$0.10 & \CM & \CM & \CM & \CM & \CM &     &     \\  
5  & NGC~5462    & 14:03:53.8,+54:22:10.81 & +360.9+075.3      & 1.99     & 8.51$\pm$0.02 & \CM &     & \CM & \CM & \CM &     &     \\  
6  & NGC~5447    & 14:02:30.5,+54:16:09.89 & --368.3--285.6    & 2.42     & 8.45$\pm$0.02 & \CM & \CM & \CM & \CM & \CM &     & \CM \\  
7  & NGC~5455    & 14:03:01.1,+54:14:27.97 & --099.6--388.0    & 2.19     & 8.39$\pm$0.01 & \CM & \CM & \CM & \CM & \CM &     & \CM \\  
8  & NGC~5447    & 14:02:27.8,+54:16:25.14 & --392.0--270.1    & 2.49     & 8.36$\pm$0.02 & \CM & \CM & \CM & \CM & \CM &     & \CM \\ 
9  & H1216       & 14:04:10.9,+54:25:19.20 & +509.5+264.1      & 3.56     & 8.29$\pm$0.06 & \CM & \CM & \CM & \CM & \CM & \CM &     \\ 
10 & NGC~5471    & 14:04:29.0,+54:23:48.56 & +667.9+174.1      & 3.81     & 8.16$\pm$0.02 & \CM & \CM & \CM & \CM & \CM & \CM &        
\enddata	
\tablecomments{
Properties of the 9 \ion{H}{2} regions from the M101 sample that were observed with {\it HST}/COS.
Note that 10 regions were originally proposed, but the {\it HST}/COS observations of Region 3 failed.
Column 1 lists the region numbers that we adopt in this paper, whereas Column 2 gives the regions common name from either
the New General Catalogue (NGC) or the \citet{hodge90} catalogue.
Column 3 gives the right ascension (R.A.) and declination (Decl.) in J2000 and Column 4 gives the CHAOS ID from \citet{croxall16},
which is composed of the offset R.A. and Decl., in arcseconds, from the galaxy center.
Column 5 lists the galactocentric distance as a fraction of the effective radii, $R_e$, and Column 6 lists the oxygen abundances, 
as updated by \citet{berg20}.
Columns 7--11 highlight regions with significant auroral line detections of [\ion{O}{3}] \W4363, [\ion{N}{2}] \W5755, 
[\ion{S}{3}] \W6312, [\ion{O}{2}] \W\W7230,7330, and [\ion{S}{2}] \W\W4069,4076. 
We note in columns 12 and 13 which \ion{H}{2} regions have detections of narrow, nebular \ion{He}{2} \W4686 emission
and broad Wolf-Rayet features. }
\label{tbl1}
\end{deluxetable*}


\begin{figure}
\begin{center}
	\includegraphics[width=\linewidth,trim=0mm 0mm 0mm 0mm,clip]{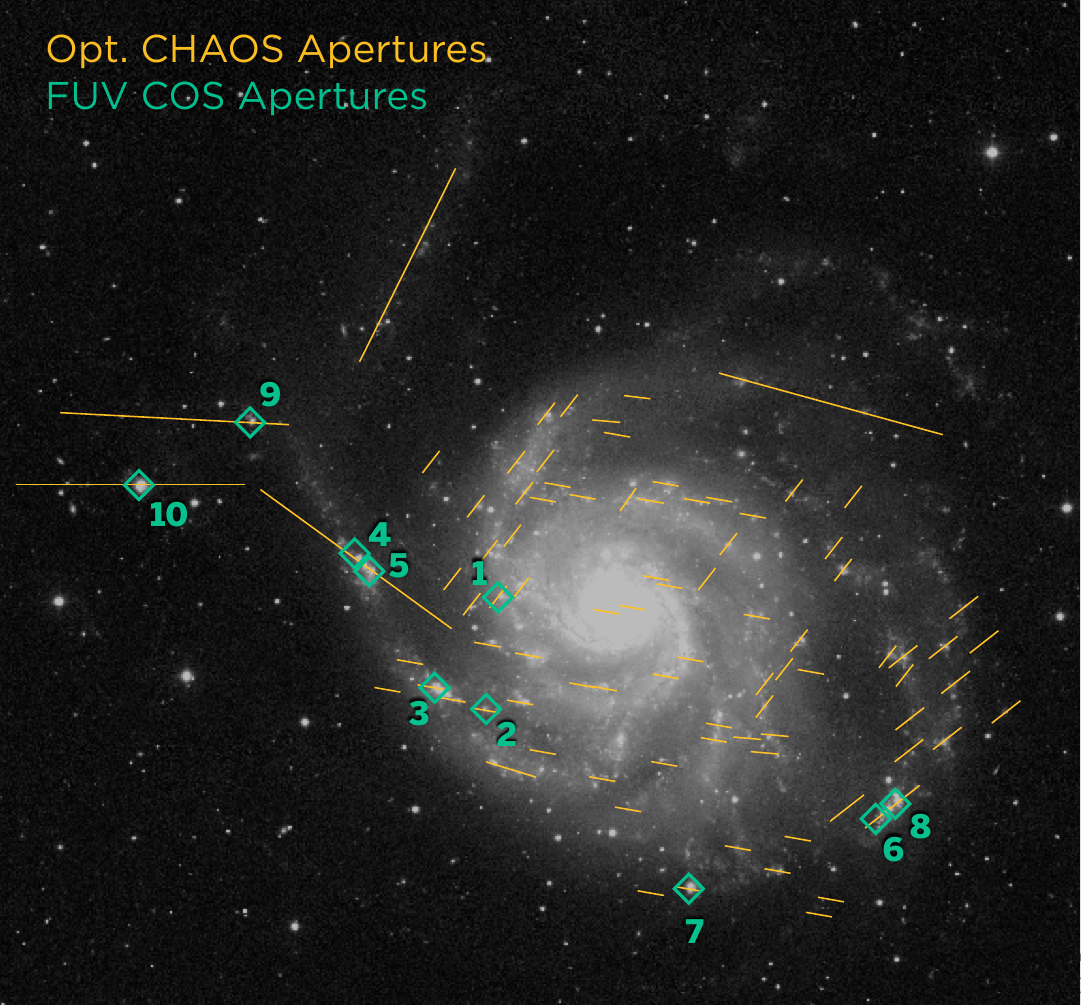}
\caption{
Image of M101 showing the \ion{H}{2} regions targeted by the CHAOS survey
using slit masks \citep[yellow lines;][]{croxall16} and the 10 regions in which 
recombination-line \ion{C}{2} \W4267 emission was detected \citep[green diamonds;][]{skillman20}.
These 10 regions were followed-up with {\it HST}/COS observations as part of the 
HST-GO-15126 program (PI: Berg). 
The regions numbers correspond to those given in Table~\ref{tbl1}. \label{fig1}}
\end{center}
\end{figure}


\section{Spectral Observations}\label{sec:obs}
\subsection{LBT/MODS Optical Spectra}
The optical observations presented here were previously reported in \citet{croxall16}
and updated in \citet{berg20}, where the interested reader can find details of 
the observations and data reduction.
Here we briefly summarize. 

Optical spectra of M101 were taken using MODS1 on the LBT during the spring semester of 2015. 
We obtained simultaneous blue and red spectra using the G400L (400 lines mm$^{-1}$, R$\approx1850$)
and G670L (250 lines mm$^{-1}$, R$\approx2300$) gratings, respectively. 
The primary objective of the CHAOS program is to detect the intrinsically weak 
temperature-sensitive auroral lines in the wavelength range from 3200--10,000 \AA\ 
in order to determine direct abundances.
Although not a design goal of the program, the observations of M101 were sensitive enough 
to detect RL emission from the \ion{C}{2} \W4267 line. 
Recombination line observations are typically made at higher spectral resolution 
than in the CHAOS program because \ion{O}{2} RLs are tightly clustered.
However, the \W4267 line is well isolated, making \ion{C}{2} RL detections and abundances 
of C$^{+2}$/H$^+$ possible from the MODS spectra.
Further discussion of the \W4267 RL detections can be found in \citet{skillman20}.

The original footprint of CHAOS observations of M101 is shown in Figure~\ref{fig1},
with the 10 regions from the current sample designated with green diamonds.
The present \ion{H}{2} region sample has a radial coverage of 
$0.2\lesssim R/R_{25}\lesssim 0.8$ and spans a range of 0.44 dex in oxygen abundance.
Properties determined for the \ion{H}{2} regions in the present sample by the CHAOS survey
are reported in \citet{croxall16}; a subset of important properties are listed in Table~\ref{tbl1}.


\begin{figure*}
\begin{center}
	\includegraphics[width=0.9\textwidth,trim=20mm 20mm 20mm 20mm,clip]{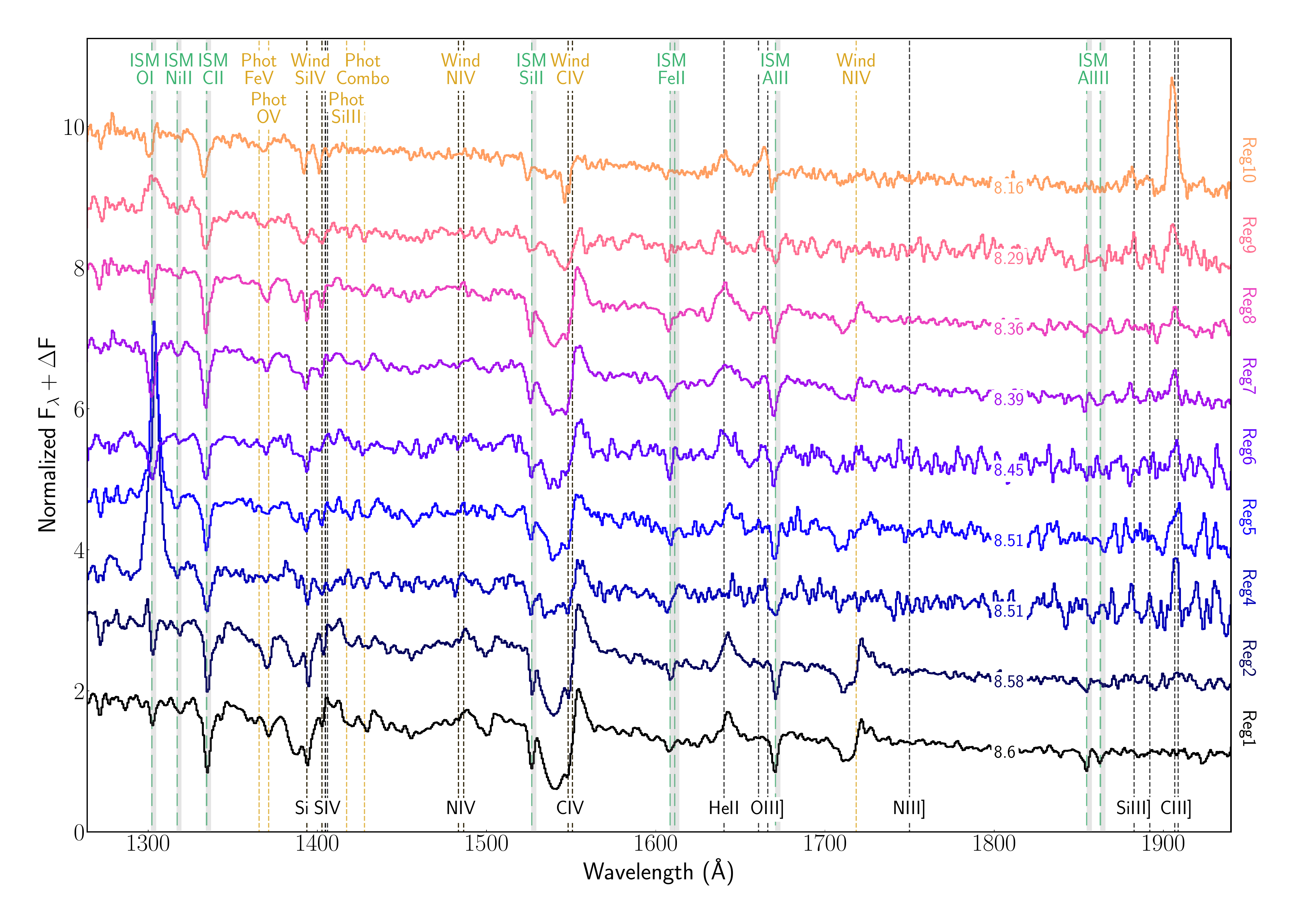}
\caption{
{\it HST}/COS G140L spectra of the 9 \ion{H}{2} regions in the CHAOS M101 sample, 
normalized at \W1800 and ordered by 
increasing gas-phase oxygen abundance from top to bottom.
The location of potential UV spectral features are denoted for ISM absorption lines (green 
long-dashed lines), stellar wind and photospheric features (gold dashed lines), and 
nebular emission lines (black dashed lines).
The combination of photospheric lines at $\sim1430$ \AA\ contains \ion{C}{3} \W\W1426,1428 and \ion{Fe}{5} \W\W1429,1430 in O stars \citep[e.g.,][]{demello00}.
Note that the resonant transitions that produce stellar wind P-Cygni profiles 
can also have ISM absorption components.
Due to the low-resolution of the G140L spectrum, MW features (locations indicated by gray 
bands) are not resolved and may contaminate other features.
Note that there is also Geocoronal \ion{O}{1} emission near \W1300 in the spectra 
of Regions 4 and 5.
\label{fig2}}
\end{center}
\end{figure*}


\subsection{HST/COS UV Spectra}
In order to measure the UV stellar and nebular properties of the \ion{H}{2} regions in M101
we obtained Cycle 25 {\it HST}/COS observations (program HST-GO-15126, PI: D. Berg).
\citet{berg16} demonstrated the utility of simultaneously observing the \ion{O}{3}] \W\W1661,1666  
and \ion{C}{3}] \W\W1907,1909 nebular emission lines in the FUV afforded by the large wavelength coverage 
(1282--2148 \AA\ in Segment A) of the low-resolution G140L grating.
Adopting this observing strategy, we observed each of 9\footnote{All 10 \ion{H}{2} 
regions in the \citet{skillman20} sample were proposed for in program HST-GO-15126, 
however, the target acquisition failed on the final target (region 3, NGC~5461). 
Since Space Telescope Science Institute deems a program as successful when 90\%\ complete,
the observation was not repeated.} 
\ion{H}{2} regions in M101 for 1-3 orbits.

Due to the proximity of M101, the large \ion{H}{2} regions fill the COS aperture; however, 
target image acquisition was efficiently achieved using a pointing offset from a nearby bright star. 
We used deep B-band LBT Large Bincoluar Cameras (LBCs) imaging from
the OSU Monitor program \citep[e.g.,][]{neustadt21}
to select bright, isolated stars near our target \ion{H}{2} regions.
The excellent astrometry of these images provided precise positions of the offset stars and 
allowed us to accurately determine offsets.
We used the ACQ/IMAGE mode with the PSA aperture and MirrorB (due to the 
brightness of the offset stars) with the COS/NUV configuration to acquire the
stars and then blind offset to the target \ion{H}{2} regions.

COS FUV spectral observations were taken in the TIME-TAG mode using the 2.5\arcsec\ 
PSA aperture and the G140L grating at a central wavelength of 1280\AA. 
We used the FP-POS = ALL setting, which takes four images offset from one another in 
the dispersion direction, increasing the cumulative S/N and mitigating the effects 
of fixed pattern noise. 
The four positions allow a flat to be created and for the spectrum to fall on different 
areas of the detector to minimize the effects of small scale fixed pattern noise.
Spectra from the 9 successful observing visits were processed with CALCOS version 3.3.4.
In order to gain signal-to-noise, we followed \citet{berg19a} and re-binned the 
spectra by a factor of six in the dispersion direction and smoothed with a 3-pixel
box-car kernel.
Nominally, the COS has a resolution of $R\sim2000$ for point sources, but this 
resolution is degraded for nebulous regions that fill the aperture.
By measuring individual photosphere absorption lines in our spectra, 
we found a typical FWHM$\sim300$ km s$^{-1}$, which corresponds to $R\sim1000$.


\begin{deluxetable*}{cccccccccccRR}
\tabletypesize{\scriptsize}
\tablecaption{Comparison of Stellar and Gas Properties}
\tablehead{
\CH{} & \CH{} & \multicolumn{6}{c}{Stellar Properties}               && \multicolumn{2}{c}{Gas Properties} & \CH{} & \CH{} \\ \cline{3-8} \cline{10-11} \\ [-3ex]
\CH{Reg.} & \CH{}   & \CH{E(B-V)$_{\rm SB99}$} & \CH{E(B-V)$_{\rm BPASS}$} & \CH{Age$_{\rm SB99}$} & \CH{Age$_{\rm BPASS}$} & \CH{Z$_{\star,{\rm SB99}}$}& \CH{Z$_{\star,{\rm BPASS}}$}  && \CH{O/H$_{neb.}$} & \CH{N/O$_{neb.}$} & \CH{$\Delta$Z$_{\rm SB99}$} & \CH{$\Delta$Z$_{\rm BPASS}$} \\ [-3ex]
\CH{\#}   & \CH{WR} & \CH{(mag.)} & \CH{(mag.)} & \CH{(Myr)} & \CH{(Myr)} & \CH{(Z$_{\odot}$)} & \CH{(Z$_{\odot}$)} && \CH{(O/H$_\odot$)}& \CH{(N/O$_\odot$)} & \CH{(Z$_{\odot}$)} & \CH{(Z$_{\odot}$)}}
\startdata
1  & \CM & 0.218$\pm$0.003 & 0.178$\pm$0.004 & 1.94$\pm$0.03 & 3.65$\pm$0.56 & 1.00$\pm$0.03 & 0.88$\pm$0.01 && 0.81$\pm$0.01 & 0.81$\pm$0.01 & 0.19 & 0.06 \\ 
2  & \CM & 0.197$\pm$0.004 & 0.172$\pm$0.003 & 1.60$\pm$0.04 & 2.76$\pm$0.19 & 1.00$\pm$0.03 & 0.88$\pm$0.01 && 0.78$\pm$0.01 & 0.74$\pm$0.01 & 0.22 & 0.10 \\ 
4  &     & 0.239$\pm$0.035 & 0.258$\pm$0.020 & 4.12$\pm$2.33 & 4.15$\pm$2.38 & 0.49$\pm$0.13 & 0.51$\pm$0.07 && 0.66$\pm$0.02 & 0.44$\pm$0.04 & -0.17& -0.15\\
5  &     & 0.245$\pm$0.014 & 0.251$\pm$0.015 & 2.42$\pm$0.64 & 5.57$\pm$4.03 & 0.89$\pm$0.11 & 0.47$\pm$0.04 && 0.66$\pm$0.01 & 0.43$\pm$0.01 & 0.23 & -0.19\\
6  & \CM & 0.311$\pm$0.021 & 0.311$\pm$0.016 & 1.64$\pm$0.77 & 1.83$\pm$2.06 & 0.93$\pm$0.15 & 0.60$\pm$0.03 && 0.58$\pm$0.01 & 0.55$\pm$0.01 & 0.35 &-0.03\\  
7  & \CM & 0.195$\pm$0.005 & 0.182$\pm$0.005 & 2.42$\pm$0.11 & 4.30$\pm$1.00 & 0.78$\pm$0.05 & 0.52$\pm$0.02 && 0.50$\pm$0.01 & 0.55$\pm$0.01 & 0.28 & 0.02 \\  
8  & \CM & 0.174$\pm$0.005 & 0.159$\pm$0.005 & 2.26$\pm$0.09 & 2.40$\pm$0.96 & 0.99$\pm$0.04 & 0.55$\pm$0.02 && 0.47$\pm$0.01 & 0.59$\pm$0.01 & 0.52 & 0.08 \\  
9  &     & 0.169$\pm$0.011 & 0.192$\pm$0.011 & 3.31$\pm$2.16 & 7.14$\pm$3.40 & 0.62$\pm$0.15 & 0.28$\pm$0.07 && 0.40$\pm$0.01 & 0.33$\pm$0.03 & 0.22 &-0.12 \\
10 &     & 0.158$\pm$0.016 & 0.150$\pm$0.007 & 9.69$\pm$3.90 & 12.2$\pm$3.05 & 0.29$\pm$0.06 & 0.22$\pm$0.02 && 0.30$\pm$0.01 & 0.34$\pm$0.01 &-0.01 &-0.08  
\enddata	
\tablecomments{Stellar and nebular properties for the nine \ion{H}{2} regions studied here.
The stellar properties are derived from the luminosity-weighted stellar continuum fits
described in Section~\ref{sec:contfit}, where columns 3--8 correspond to the derived 
reddening, age, and metallicity of the massive star population from both \texttt{SB99} and \texttt{BPASS} fits.
The differences between the stellar model metallicities and the gas-phase metallicity are listed in
columns 9--10.
For comparison, regions with identified WR features are noted in Column 2, while the
nebular O/H and N/O abundances are given in columns 9--10.
All abundances are relative to the solar scale from \citet{asplund21}, e.g., N/O$_{neb.}=0.813$ N/O$_\odot$
is equivalent to 81.3\%\ the solar N/O value.}
\label{tbl2}
\end{deluxetable*}


Figure~\ref{fig2} shows the FUV spectra for the 9 \ion{H}{2} regions observed.
The observed wavelength coverage of G140L is $\sim900-1195$ for Segment B 
and $\sim1282-2000$ for Segment A\footnote{Segment A coverage is quoted out to 2148 \AA, however, 
previous experience with G140L spectra suggest G140L 
Segment A spectra only have utility out to $\sim2000$ \AA.}. 
The 105 \AA\ gap between Segments A and B mitigates the risk of strong Geocoronal Ly$\alpha$ 
emission on the detector, however, this also causes our spectra to miss the \ion{N}{5}
\W\W1238,1242 stellar wind feature. 
The spectra were corrected for the redshift of M101 ($z=0.0008$),
but not for the small offsets resulting from the galactocentric rotational velocity 
of a given \ion{H}{2} region.

In order to allow a close comparison of feature trends, the spectra in Figure~\ref{fig2} are 
vertically offset from one another and ordered from top to bottom by increasing metallicity.
Nebular features are labeled by gray short-dashed lines, while stellar wind and photospheric
features are gold, and ISM absorption is labeled green.
Some features are complex combinations of multiple components, such as \ion{C}{4},
which can have stellar, nebular, ISM components, and, at the small redshift of M101,
contamination from Milky Way absorption.
Some features are clearly sensitive to metallicity, such as the \ion{S}{4}, \ion{C}{4}, and \ion{N}{4}
stellar wind features, which strengthen with increasing metallicity, and the \ion{C}{3}]
nebular emission, which generally weakens in strength at higher metallicity.
We also note the presence of \ion{He}{2} emission, which can be both stellar
and nebular in origin.
As we will see later, the strong, broad \ion{He}{2} features in regions 1, 2, 6, 7, and 8 are 
likely stellar, as strong Wolf-Rayet (WR) features are noted in their optical spectra.
On the other hand, narrow \ion{He}{2} emission was found in the optical spectra of
regions 9 and 10, so their UV \ion{He}{2} features may be nebular (see Table~1).


\section{Stellar Continuum Fitting}\label{sec:contfit}


\begin{figure*}
\begin{center}
  \includegraphics[width=0.99\textwidth, trim=60mm 50mm 60mm 60mm, clip]{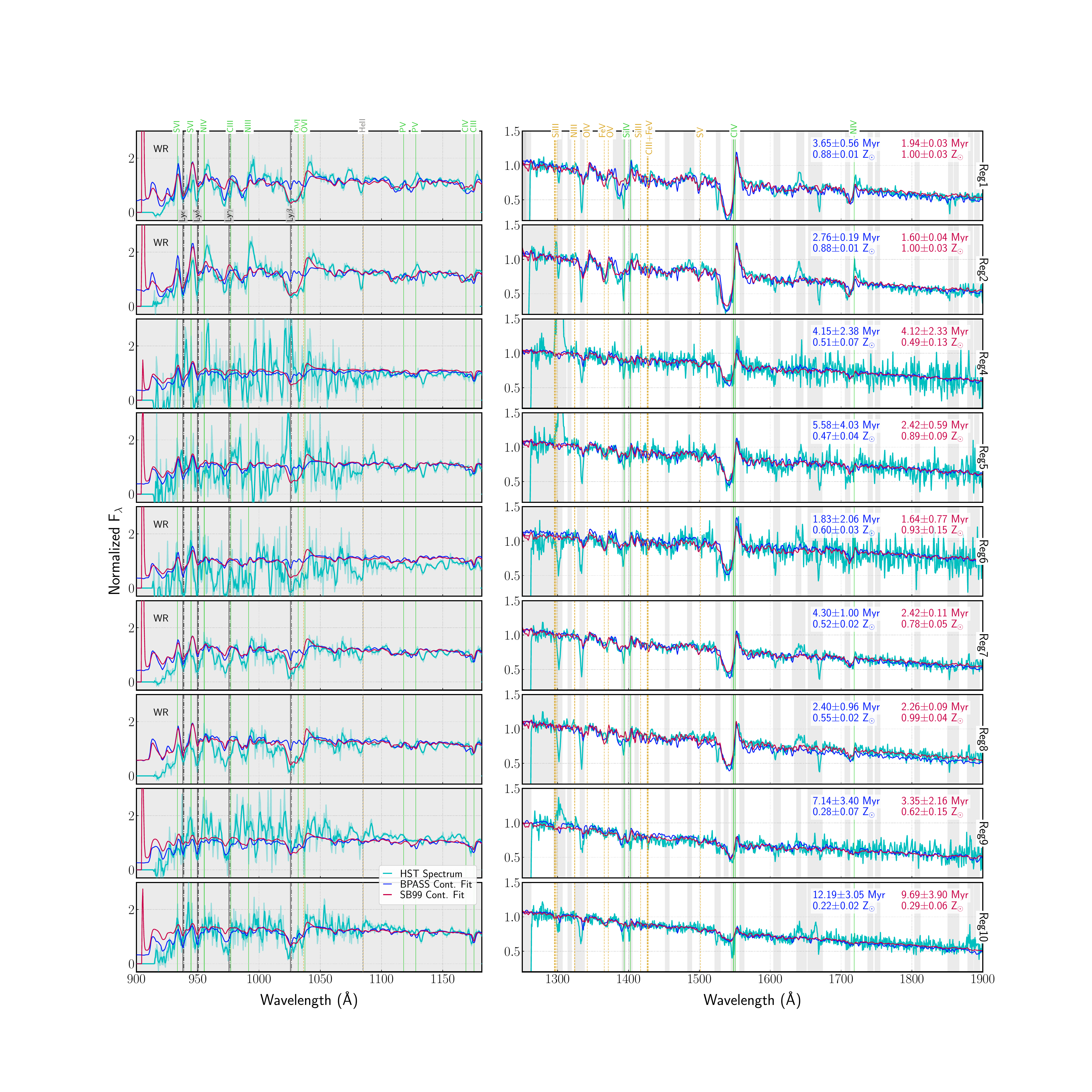}  
\caption{Luminosity-weighted best fit stellar models to the 
{\it HST}/COS FUV spectra of the M101 \ion{H}{2} region sample.
Observed spectra are shown in turquoise, with the best-fit stellar \texttt{Starburst99} 
and \texttt{BPASS} models overlaid in red and blue, respectively.
Rest-wavelengths of stellar wind and photospheric features are shown by 
solid green and dashed gold lines, respectively. 
Additionally, H and He lines are denoted by solid gray lines; the MW H lines are
indicated with dashed-dotted gray lines.
Fits were performed after masking any contamination from MW sky lines,
ISM absorption features, and nebular emission lines (gray-shaded bands).
The resulting age and metallicity that best characterizes the ionizing 
stellar population is labeled in the upper right hand corner
of each panel for both the \texttt{Starburst99} (red) and \texttt{BPASS}
(blue) models. 
We also highlight the excellent \texttt{PoWR} WR model fit to the 
\ion{N}{3} \W991 feature of Region 1 in the yellow inset box.
In general, there is a progression from strong stellar features at the
highest metallicities (top) to small features at the lowest 
metallicities (bottom). \label{fig3}}
\end{center}
\end{figure*}


The interplay between gas and stars is one of the most fundamental, 
yet unsettled, drivers of galactic baryon evolution.
The FUV ($\sim912-2000$ \AA) is arguably the richest spectral regime in 
diagnostic features characterizing the processes involved in the baryon feedback cycle.
This spectral range not only provides access to the only strong C nebular emission lines, 
but also reveals stellar wind and photospheric features that are more sensitive to
the age and metallicity of the ionizing stellar population than any optical
diagnostic features \citep[e.g.,][]{chisholm19}. 
Consequently, the FUV continuum offers a valuable means to characterize the stellar clusters 
within \ion{H}{2} regions, enabling a direct link between the properties of massive stars 
and the nebular emission lines they power. 
Further, in the high surface-brightness \ion{H}{2} regions targeted here, the 
ionizing stellar population recently formed from the surrounding gas such that 
the nebular and stellar chemical compositions are expected to be similar.
Here we take advantage of the {\it HST}/COS FUV spectra of M101 to model the 
stellar continua of our \ion{H}{2} region sample and better understand the 
conditions powering their observed emission-line spectra.


\subsection{Fitting Method}


We fit the stellar continuum following the procedure detailed in \citet{chisholm19}.
This method uses stellar population synthesis modeling to characterize 
the ages and metallicities of the ionizing stellar populations.  
Fitting was performed as a linear combination of single-age, fully theoretical 
stellar continuum burst models matched to the non-ionizing FUV continua of
our sample.
We use two sets of theoretical stellar models:
(1) single star models from \texttt{Starburst99} \citep[SB99;][]{leitherer99} and
(2) binary star models from \texttt{BPASS} v2.2.1 
\citep[][]{eldridge17, stanway18}. 
With both set of models we employ a range of possible metallicities covering 
0.05, 0.20, 0.40, 1.0, and 2.0 Z$_\odot$ and a range of burst ages (1--40 Myr), 
use a high-mass cutoff of 100 $M_\odot$.
The best intrinsic spectrum is combined with a foreground dust screen model and 
the \citet{reddy16} attenuation law to determine the reddening due to dust ($E(B-V)$).

We fit the entire observable range of the {\it HST}/COS G140L segment A spectra, covering $\sim1300-2000$ \AA.
Next, we convolved the 0.4 \AA\ resolution of the \texttt{SB99} models and the 
1 \AA\ resolution of the \texttt{BPASS} models to the spectral resolution of the 
individual \ion{H}{2} region spectra \citep[e.g.,][]{meynet94, leitherer99}. 
In order to isolate the purely-stellar portions of the spectra, we mask any 
features that are not produced by the stars, including Milky Way absorption 
features and ISM emission and absorption lines from the \ion{H}{2} regions.
A list of features considered for masking can be found in \cite{leitherer11}.
The errors on the stellar fits were determined using a boot-strap Monte
Carlo procedure, where 300 iterations were performed 
\citep[the number of iterations needed to reach convergence for spectra with these S/N ratios;][]{chisholm19}
using a modified version of the observed spectrum with error randomly sampled 
from a normal distribution. 
Further details of the fitting method can be found in \citet{chisholm19}.

The stellar continuum modeling simultaneously constrains the light-weighted age, 
metallicity, and reddening ($E(B-V)$) of the ionizing massive star 
population, the results of which are listed in Table~\ref{tbl2}.
We note that Regions 1 and 2 were best fit by \texttt{SB99} with super-solar light-weighted 
metallicities that are significantly higher than their gas-phase abundances.
Therefore, we reran the \texttt{SB99} fits without the 2.0 Z$_\odot$ model, 
effectively enforcing an upper metallicity limit of solar for these two \ion{H}{2}
regions in order to more closely match the observed gas-phase oxygen abundance range.

The uncertainties reported from the Monte Carlo procedure for the reddening, age,
and metallicity of the stellar population are listed in Table~\ref{tbl2}.
In general, the uncertainties on the reddening and metallicity values are small 
($\lesssim25$\%\ for \texttt{SB99} and $\lesssim10$\%\ for \texttt{BPASS}).
However, a few of the regions have large age uncertainties, especially the regions
lacking WR features and regions older than 5 Myr.
These more uncertain ages are likely due to degeneracies arising from the lack of 
strong spectral features able to distinguish older population ages.
Similarly, \citet{sirressi22} used integrated stellar population spectra with the 
same continuum-fitting method presented here and compared to resolved star photometry 
to show that ages could be measured with high accuracy ($<1$ Myr) 
for populations with ages between 2 and 5 Myr.
This suggests that our youngest stellar ages may have underestimated uncertainties,
but they are robust within 1 Myr.

\subsection{\texttt{SB99} and \texttt{BPASS} Comparison}
The best stellar continuum fits using both \texttt{SB99} and \texttt{BPASS} 
are shown in Figure~\ref{fig3}, over-plotted on the observed {\it HST} 
spectra, where the masked portions for each \ion{H}{2} region are denoted by 
gray bands.
However, the \texttt{SB99} and \texttt{BPASS} stellar continuum fits look very visually
similar for most of the \ion{H}{2} regions over the fit range (right column panels).
Comparing the results in Table~\ref{tbl2}, we find that the \texttt{BPASS}
fits have older light-weighted ages than the \texttt{SB99} models by 1.6 Myr
on average.
Countering this effect, on average, the \texttt{BPASS} fits have 23\% lower 
metallicities than the \texttt{SB99} fits.
The fitted reddening values are similar between \texttt{SB99} and \texttt{BPASS}, 
with an average difference of only 0.006. 

A visual comparison of the stellar features in the \texttt{SB99} and \texttt{BPASS}
models relative to the observed spectra can provide further insight.
In the fitted part of the spectra (right column panels), there are three stellar 
wind features -- \ion{Si}{4} \W\W1393,1403, \ion{C}{4} \W\W1548,1550, and
\ion{N}{4} \W1720 -- whose wind profiles are all fit reasonably well by both 
\texttt{SB99} and \texttt{BPASS}.
However, excess \ion{Si}{4} and \ion{N}{4} emission are seen in several
of the WR regions. 
Additionally, neither \texttt{SB99} nor \texttt{BPASS} have WR prescriptions that
can produce sufficiently broad \ion{He}{2} emission in these regions. 

While not fit with the stellar models, we also examined the bluest portion of 
the COS spectra.
The $\sim900-1200$ \AA\ portion of the FUV is notoriously difficult to fit due to
the densely packed suite of stellar and ISM lines in the observed galaxy and
low-ionization resonance lines formed in the intervening gas, including numerous
Lyman series H lines and Lyman and Werner bands  of H$_2$ \citep[e.g.,][]{tumlinson02}.
This leaves little to no uncontaminated continuum to fit \citep[see, e.g.,][]{willis04}.
Therefore, it is interesting that both the 
\texttt{SB99} and \texttt{BPASS} models produce good overall fits to the observed spectra, 
despite being unconstrained by the data in this challenging wavelength regime.

On the other hand, the fit of the \texttt{SB99} models to the most prominent features
in the bluest spectral regime is rather impressive.
In particular, the \texttt{SB99} models trace and confirm the \ion{O}{6} \W\W1032,1038 
P-Cygni profile in Regions 1, 2, 5, 6, 7, and 8.
The \ion{O}{6} feature is further evidence of very young stars in these regions ($<3$ Myr; see next 
section), favoring the younger ages of the \texttt{SB99} fits.
Note, however, that the \texttt{BPASS} models do not incorporate X-ray winds, 
and so they will not be able to fit the \ion{O}{6} (I.P. $\gtrsim 114$ eV) feature.
It would also be useful to fit the age-sensitive \ion{N}{5} \W\W1238,1242 feature in
order to confirm the young ages of the massive star population, but \ion{N}{5} 
unfortunately falls in the G140L gap (implemented to avoid Geocoronal Ly$\alpha$) at 
the low redshift of M101. 


\begin{figure}
\begin{center}
	\includegraphics[width=0.475\textwidth]{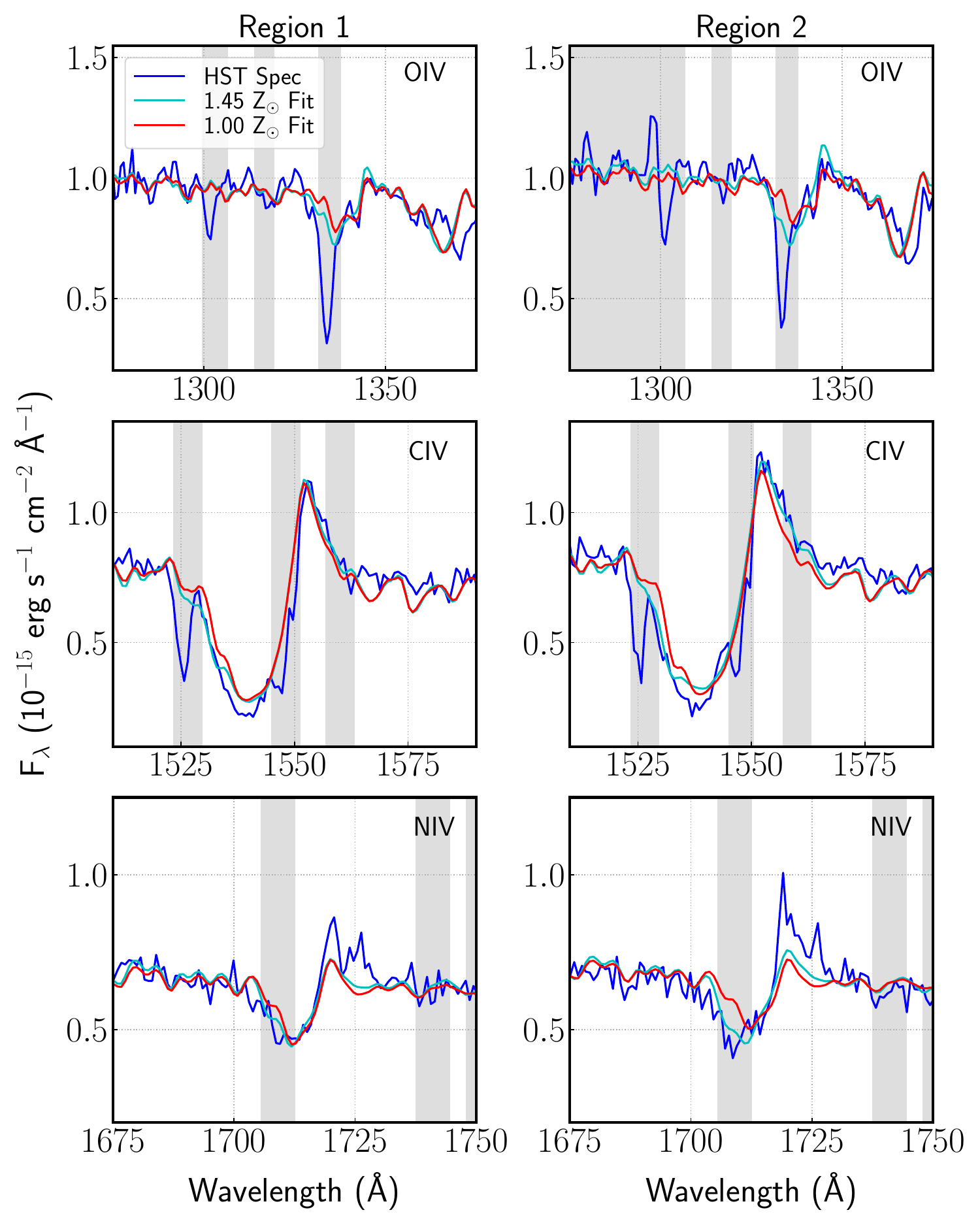}
\caption{Comparison of the \ion{O}{4}, \ion{C}{4}, and \ion{N}{4} stellar features in
Regions 1 and 2 for \texttt{Starburst} models with different light-weighted metallicities.
Regions masked for possible contamination from MW sky lines or
ISM absorption features are indicated by gray-shaded bands.
There is very little difference between the two sets of models, and the models 
fail to reproduce the peak of the emission in the \ion{N}{4} P-cygni profile,
demonstrating the need for both improved models and higher resolution and S/N spectra. \label{fig4}}
\end{center}
\end{figure}

 
\subsection{Single- and Mutli-Population Comparison}

\begin{figure}
\begin{center}
	\includegraphics[width=0.45\textwidth, trim=0mm 0mm 5mm 0mm,clip]{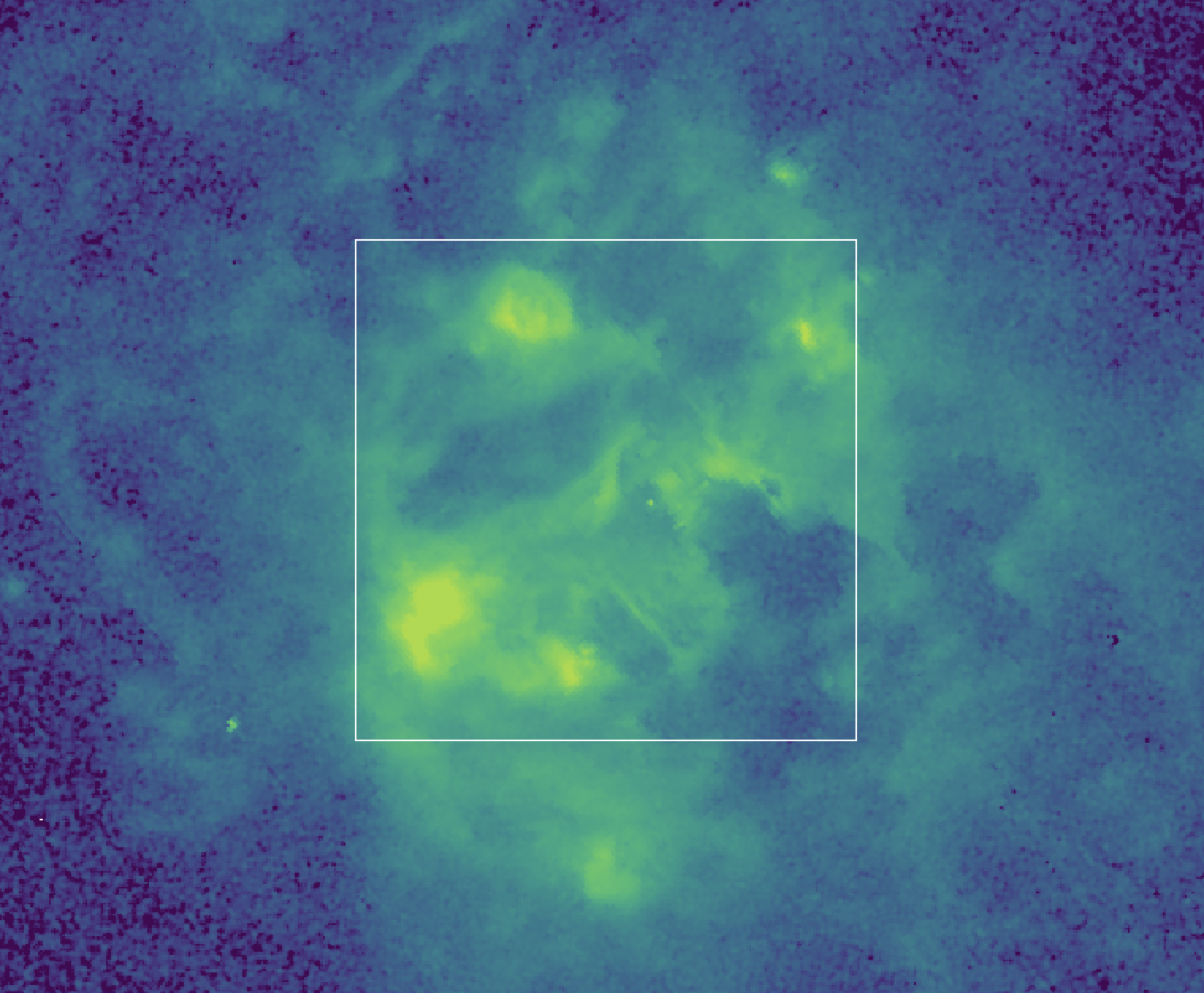}
\caption{This image of Region 10 (NGC 5471) using the F656N filter on HST/WFPC2 
(PID: HST-GO-6829; PI: You-Hua Chu)
shows multiple knots of H$\alpha$ emission composing the complex, giant \ion{H}{2} region.
This suggests that multiple star associations are contributing to the gas ionization.
For reference, the white box spans 10\arcsec$\times$10\arcsec. \label{fig5}}
\end{center}
\end{figure}

Typically, simple stellar populations (SSPs) assuming a single-burst (single-population) 
are used to characterize the stellar continuum of an \ion{H}{2} region, while multiple 
single-burst models (multi-population) are combined to characterize the integrated spectra
of galaxies. 
As demonstrated for Region 10 (NGC~5471) in Figure~\ref{fig5}, 
many of the \ion{H}{2} regions in M101 are giant complexes containing multiple knots of 
star-formation and earning them their own NGC classifications. 
For example, \citet{garcia-benito11} performed a photometric study of the star formation 
history of NGC~5471 showing that star formation has been more or less continuous for
at least 100 Myr.
Further, they found that both the integrated photometry and resolved-star color-magnitude 
diagram of NGC~5471 is consistent with star formation of two different ages.

More recently, \citet{sirressi22} used Legacy ExtraGalactic UV Survey (LEGUS) HST multiband 
photometry and CLusters in the UV as EngineS (CLUES) FUV HST/COS spectroscopy of 20 
young stellar clusters hosted in 11 nearby star-forming galaxies to examine their 
stellar populations. 
Using single-population fits, two-population fits, and the same multi-population fit method 
used here, \citet{sirressi22} found that most of the star-forming regions examined 
(16/20) are populated by two or more stellar populations.
In fact, a second population was required for these regions to accurately reproduce the 
observed level of spectral continuum and the P-Cygni line profiles simultaneously.
Given the complicated nature of the giant \ion{H}{2} regions in M101 and their complex
star formation histories, we adopted the multi-population method to characterize their 
stellar continua in our analysis above.

\begin{figure*}
\begin{center}
	\includegraphics[width=1.0\textwidth, trim=30mm 0mm 30mm 0mm,clip]{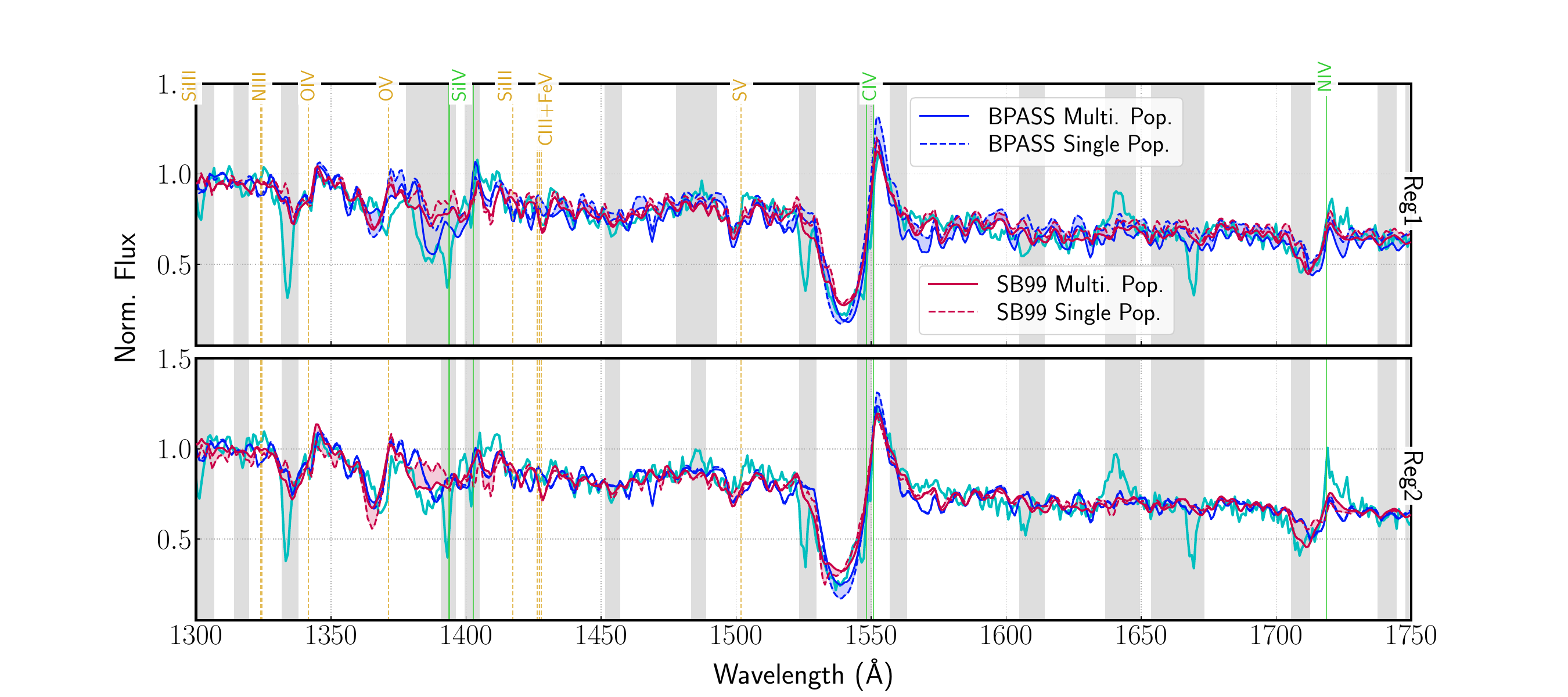}
\caption{Comparison of the single-population best-fit stellar models (dashed lines)
to the luminosity-weighted multi-population best-fit stellar models for Region 1 
(top panel) and Region 2 (bottom panel).
Observed spectra are shown in turquoise, with the best-fit stellar continuum models of 
\texttt{Starburst99} (red) and \texttt{BPASS} (blue) overlaid.
Differences between the single- and multi-population fits are emphasized by red and blue 
shading for the \texttt{Starburst99} and \texttt{BPASS} models, respectively.
(blue) models. 
As in Figure~\ref{fig3}, stellar wind (solid green lines), photospheric features 
(dashed gold lines), and masked regions (gray-shaded bands) are shown.
Overall, the multi-population fits better reproduce the observed stellar wind features,
and are adopted for the analysis in this work.
\label{fig6}}
\end{center}
\end{figure*}

To test the validity of our multi-population assumption, we also fit the two regions showing 
the strongest stellar features (Regions 1 and 2) with single-population fits.
The resulting single-population \texttt{SB99} and \texttt{BPASS} fits (dashed-lines) are shown in Figure~\ref{fig6} 
in comparison to their corresponding multi-population fits (solid lines).
In general, the multi-population fits better fit the stellar wind profiles.
For Region 1, the single- and multi-population \texttt{SB99} fits show little difference, but the 
multi-population \texttt{BPASS} fit more closely matches the \ion{Si}{4} \W\W1393,1403 and \ion{C}{4} 
\W\W 1548,1500 outflow profiles than the single-component \texttt{BPASS} fit.
For Region 2, both the \texttt{SB99} and \texttt{BPASS} fits show differences between the single- and 
multi-population fits.
In particular, the multi-population fits better reproduce the \ion{Si}{4} \W\W1393,1403, 
\ion{C}{4} \W\W 1548,1500, and the \ion{N}{4} \W1720 profiles.

The resulting stellar continuum properties for the single- and multi-population \texttt{SB99} and 
\texttt{BPASS} fits are compared in Table~\ref{tbl2b}.
In general, the reduced $\chi^2$ values are similar between the single- and multi-component 
fits, but are smaller for the multi-component fits. 
Additionally, the multi-component fit properties have much smaller uncertainties relative 
to the single-component fits. 
However, an important aspect of the multi-population \texttt{BPASS} fits is that they allow for
slightly older ages.
We, therefore, conclude that there are not significant differences between the single-population
and multi-population fits, but that the multi-population fits provide a small improvement
in the complex \ion{H}{2} regions of M101.


\begin{deluxetable*}{rCCCCCCC}
\tabletypesize{\small}
\tablecaption{ Comparison of Single- vs. Multi-Population Stellar Continuum Properties}
\tablehead{
\CH{} & \multicolumn{3}{c}{Reg 1} && \multicolumn{3}{c}{Reg 2} \\ \cline{2-4} \cline{6-8} \\ [-2.5ex]
\CH{}           & \CH{Single} & \CH{Multi.} & \CH{}         && \CH{Single} & \CH{Multi.} & \CH{} \\ [-2.5ex]
\CH{Property}   & \CH{Pop.}   & \CH{Pop.}   & \CH{$\Delta$} && \CH{Pop.}   & \CH{Pop.}   & \CH{$\Delta$} }
\startdata
Z$_{\star,{\rm SB99}}$ (Z$_\odot$)  & 1.000\pm0.238 & 1.000\pm0.028 & +0.000 && 1.000\pm0.238 & 1.000\pm0.030 & +0.000\\
Z$_{\star,{\rm BPASS}}$ (Z$_\odot$) & 1.000\pm0.238 & 0.876\pm0.009 & -0.124 && 1.000\pm0.238 & 0.881\pm0.006 & -0.119\\
Age$_{\rm SB99}$ (Myr)              & 2.000\pm2.000 & 1.940\pm0.033 & -0.060 && 1.000\pm2.000 & 1.600\pm0.040 & +0.600\\
Age$_{\rm BPASS}$ (Myr)             & 1.000\pm2.000 & 3.650\pm0.558 & +2.650 && 1.000\pm4.000 & 2.758\pm0.192 & +1.758\\
E(B-V)$_{\rm SB99}$ (mag.)          & 0.247\pm0.033 & 0.218\pm0.003 & -0.029 && 0.208\pm0.028 & 0.197\pm0.004 & -0.011\\
E(B-V)$_{\rm BPASS}$ (mag.)         & 0.223\pm0.021 & 0.178\pm0.004 & -0.045 && 0.180\pm0.010 & 0.172\pm0.004 & -0.008\\
Reduced $\chi^2_{\rm SB99}$         & 14.63         & 14.22         & -0.41  && 8.92          & 8.89          & -0.03  \\
Reduced $\chi^2_{\rm BPASS}$        & 9.72          & 8.27          & -1.45  && 5.86          & 5.73          & -0.13
\enddata	
\tablecomments{
Comparison of the properties derived for the single- and multi-population \texttt{SB99} and \texttt{BPASS} 
stellar continuum fits for Regions 1 and 2.
The single-population fits tend to drive metallicities higher and ages lower relative
to the multi-population fits.
The multi-population fits also tend to have smaller reduced $\chi^2$ values.}
\label{tbl2b}
\end{deluxetable*}


\section{Characteristics of The FUV Spectra of M101}\label{sec:spectra}
\subsection{Stellar Continuum Properties}\label{sec:continuum}
Figure~\ref{fig3} shows stellar-wind P-Cygni profiles of \ion{O}{6} \W\W1032,1038, 
\ion{Si}{4} \W\W1393,1402, \ion{C}{4} \W\W1548,1550, and \ion{N}{4} \W1720.
Since more massive and luminous stars have shorter main sequence lifetimes, the
P-Cygni profile strengths constrain the age of the ionizing stellar population.
Higher ionization state profiles, such as \ion{O}{6} and \ion{N}{5}, are 
strongest in stars with lifetimes of 2--3 Myr, while lower ionization state 
profiles, such as \ion{C}{4} and \ion{Si}{4}, are more prominent in stars with 
lifetimes near 5 Myr. 
This trend can be seen in Figure~\ref{fig3}, were the \ion{C}{4} profile is generally
stronger for the lowest stellar population ages and weak for the highest ages. 
This is further evidenced by Regions 1, 2, 6, and 7 having the strongest \ion{O}{6} P-Cygni 
profiles and the youngest \texttt{SB99} ages. 
Note, however, the \ion{O}{6} wind profile is contaminated by strong Ly$\beta$ 
absorption or emission, and so requires careful evaluation. 

P-Cygni stellar-wind profiles are also sensitive to the metallicity of the stellar 
photospheres, where continuum photons are absorbed and drive gas from the stellar surface.
However, this metallicity sensitivity declines above solar.
In Figure~\ref{fig4}, we show the \ion{O}{4}, \ion{C}{4}, and \ion{N}{4} stellar features
for \ion{H}{2} regions 1 and 2 in our sample, both of which were initially best fit with a 
super-solar luminosity-weighted stellar continuum metallicity.
In comparison, we show their respective adopted best-fit $Z=Z_\odot$ stellar continuum 
models (red) and a super-solar $Z=1.45\ Z_\odot$ model (cyan).
In both regions, the \ion{O}{4} \W1341 photospheric feature is well fit by both models 
blueward of the deep \ion{C}{2} \W1335 ISM absorption feature.
The \ion{C}{4} \W\W1548,1550 wind profiles are also well fit, but the models slightly 
underfit the depth and height of the P-Cygni profiles, regardless of metallicity.
For the \ion{N}{4} \W1719 wind feature, the absorption component of the P-Cygni profile 
is well fit blueward of the \ion{Ni}{2} ISM absorption feature, but both models fail to 
reach the strength of the emission component. 
However, an additional \ion{N}{4} source may be contributing to this profile, 
as both Regions 1 and 2 appear to have \ion{N}{4}] \W1487 emission in Figure~\ref{fig2}.

\begin{figure}
\begin{center}
	\includegraphics[scale=0.425,trim=0mm 0mm 0mm 0mm,clip]{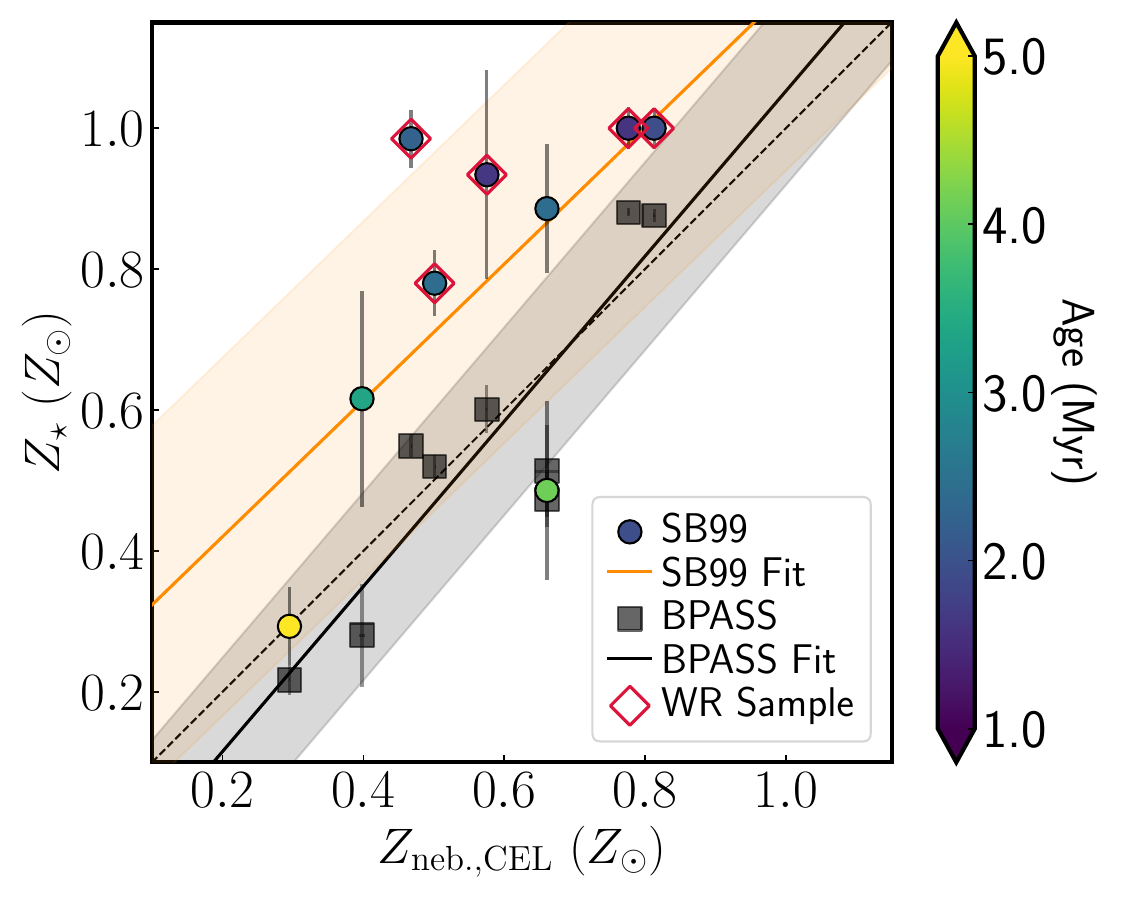}
\caption{Comparison of the derived stellar population metallicity versus
the nebular metallicity, as derived from the CEL direct method.
As expected, the stellar and nebular metallicities generally increase in lock step,
however, stellar metallicities are offset to higher values than nebular metallicities
for the youngest-aged ionizing stellar populations. 
These outliers also tend to show WR features in their optical spectra (red diamonds),
suggesting the WR phase is strongest for young ($t < 3$ Myr),
 metal-rich ($Z>0.9Z_\odot$) massive stars. \label{fig7}}
\end{center}
\end{figure}

In Figure~\ref{fig7}, we compare the derived stellar population metallicity versus
the nebular metallicity, as derived from the collisionally-excited lines direct method.
As expected, the stellar and nebular metallicities generally correlate. 
Interestingly, the \texttt{BPASS} stellar metallicities are generally consistent with the 
nebular metallicities, while the \texttt{SB99} stellar metallicities are offset to higher 
values, especially for the youngest-aged stellar populations. 
We fit a linear relationship to both the \texttt{BPASS} and \texttt{SB99} trends using the 
\texttt{Python} Bayesian linear regression code 
\texttt{linmix}\footnote{\url{https://github.com/jmeyers314/linmix}},
which is an implementation of the linear mixture model algorithm developed by 
\citet{kelly07} to fit data with uncertainties on two variables, including explicit 
treatment of intrinsic scatter. 
The \texttt{SB99} fit, shown in orange, indicates that, on average, the stellar
metallicities are offset to higher metallicities by 0.25 dex. 
Figure~\ref{fig7} also shows a cluster of \texttt{SB99} outliers at the highest stellar 
metallicities considered.

Taken as a whole, to match the strong stellar wind profiles of the youngest stellar 
populations that contain observed WR profiles (red diamonds in Figure~\ref{fig7}), 
\texttt{SB99} models require very young populations ($<3$~Myr) that are more metal-enriched 
than implied by their gas-phase metallicities. 
This may be because the lower-metallicity stellar evolution tracks do not produce significant 
amounts of evolved stars \citep{leitherer11}. 
Therefore, the code requires younger and more metal-enriched stellar populations to match 
the strong profiles in the populations with observed WR stars. 
Meanwhile, \texttt{BPASS} fits do produce sufficiently strong wind profiles that mimic the 
observed profiles at metallicities that are consistent with the observed gas-phase metallicities. 
This may be because the binary synthesis in \texttt{BPASS} allows for additional 
physical pathways to create the evolved (WR) stars observed in the spectrum \citep{eldridge17}. 
While \texttt{BPASS} does fit some of these lines better, many of the WR features are still 
poorly fit by population synthesis models. 
Thus, these stellar populations provide a strong test-bed to constrain future WR atmosphere and 
evolution models, and their incorporation to stellar population models.


\subsection{\ion{He}{2} Emission}\label{sec:HeII}
Emission from He$^{+2}$ recombination produces the \ion{He}{2} emission features commonly
observed at \W1640 in the rest-frame FUV and \W4686 in the rest-frame optical.
\ion{He}{2} emission is expected from a number of sources, including shocks, X-ray binaries,
extremely metal-poor massive stars, and the winds of massive stars, such as WR stars.
WR stars are an interesting source because they are extremely hot stars that 
have main-sequence lifetimes $\lesssim5$ Myr \citep[e.g.,][]{abbott87, crowther07} being
observed during a short-lived supergiant phase.
During this phase, WR stars produce broad \ion{He}{2} emission lines in both the optical and 
FUV due to their strong winds that increase at higher metallicity \citep[e.g.,][]{schaerer98}.

\citet{chisholm19} used light-weighted stellar populations to show that the \ion{He}{2}
profile is also very sensitive to age.
For the youngest populations in their study ($\lesssim 5$ Myr), \ion{He}{2} \W1640 
has a strong, broad profile, while the oldest populations ($> 20$ Myr) show a weak
\ion{He}{2} absorption feature.
They also found that their multiple-age fits to the FUV \ion{He}{2} \W1640 region were 
poor, especially in the presence of strong WR emission, with \texttt{SB99}  
performing worse than \texttt{BPASS}.
Similar to the results of \citet{chisholm19}, Figure~\ref{fig3} shows that our best-fit 
stellar models fail to produce the requisite number of \ion{He}{2} photons from WR stars to 
match the broad \ion{He}{2} emission observed in our {\it HST}/COS spectra of 
\ion{H}{2} regions in M101, especially for the youngest stellar populations 
\citep[see, also,][]{leitherer18}. 
Further, \citet{martins23} used \texttt{BPASS} models with an initial mass function upper
mass cutoff of 300 $M_\odot$ to show that not even the winds of very-massive stars could 
reproduce the broad strong \ion{He}{2} observed here.

\subsection{Wolf-Rayet \& Very Massive Star Features}
\subsubsection{Optical Characteristics}
\begin{figure*}
\begin{center}
	\includegraphics[width=0.975\textwidth]{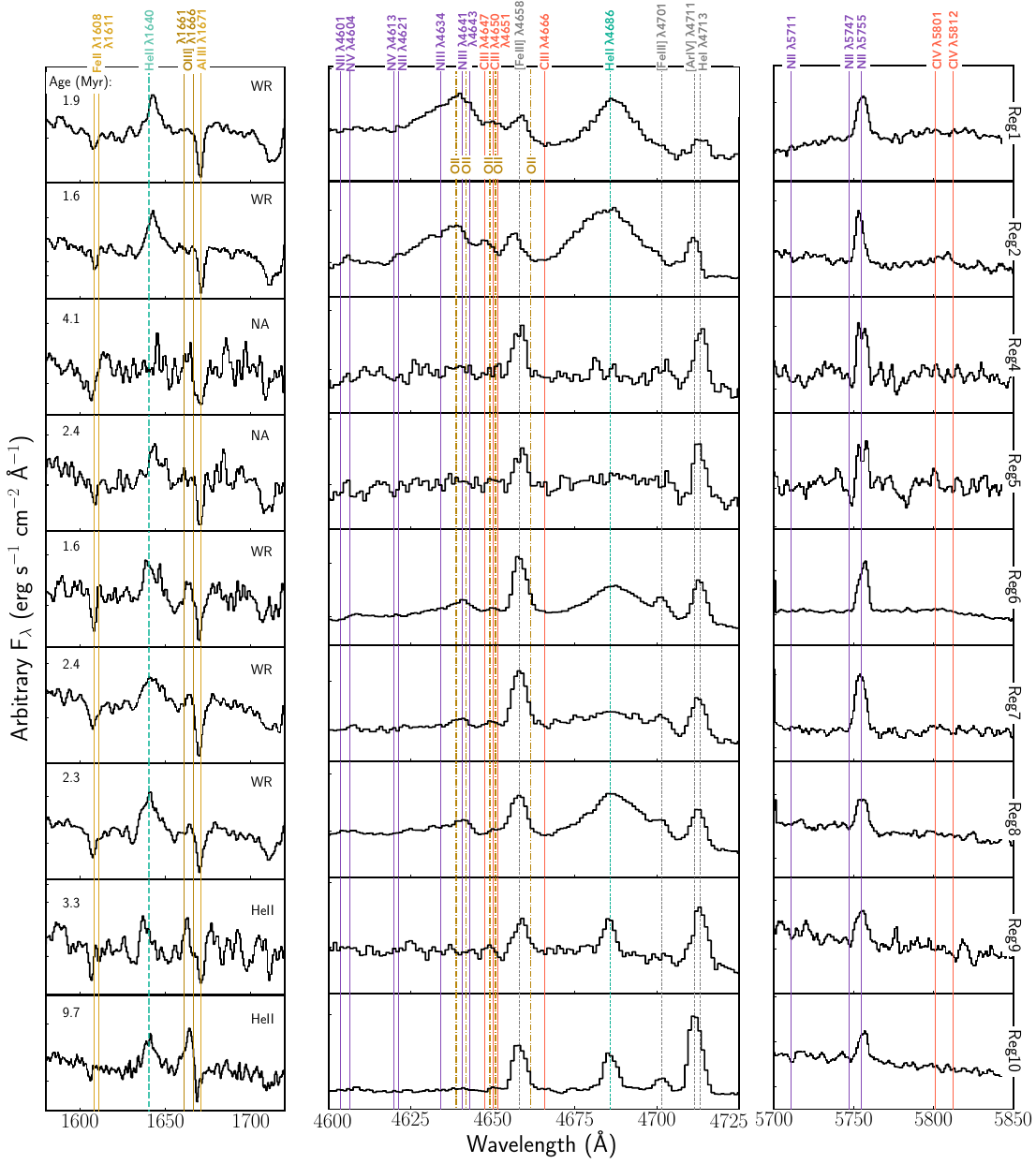}
\caption{Comparison of FUV \ion{He}{2} emission features ({\it left panel}) 
with the optical WR blue bump ({\it middle panel}) and red bump ({\it right panel}) features. 
We detect strong blue WR features in several of the regions, indicating the presence of 
nitrogen-type WR (WN) stars.
Regions identified by \citet{croxall16} to have narrow, nebular \ion{He}{2} 
emission or WR emission in their optical spectra are labeled accordingly in 
the upper right-hand corner of the left-panel plots.
We do not detect any red bumps in the right panel, and thus lack explicit evidence of 
carbon-type WR (WC) stars, but there may be \ion{C}{3} contributions to the blue bump in the middle panel. 
\label{fig8}}
\end{center}
\end{figure*}

Given the low spectral resolution of the FUV spectra ($\gtrsim 300$ km s$^{-1}$), 
the broad \ion{He}{2} emission at \W1640 is difficult to interpret.
Therefore, it is useful to look to the optical spectra ($\Delta v \sim 150$ km s$^{-1}$)
to determine whether the \ion{He}{2} emission appears narrow and nebular in origin or 
is broad due to WR or very-massive star ($>100\ M_\odot$; VMS) winds. 
In Figure~\ref{fig8}, we plot the \ion{He}{2} portion of the FUV spectra in comparison to 
the two optical regions that are commonly used to spectroscopically identify the presence 
of WR stars.
The first optical feature, known as the blue bump (middle column), is a complex of features 
near \W4650, including \ion{N}{5} \W4603, \ion{N}{3} \W\W4634,4641, \ion{C}{3} \W4647,4666, 
\ion{C}{4} \W4658, and \ion{He}{2} \W4686.
The second feature is a blend of \ion{C}{4} transitions at \W\W5801,5812, known as the red 
bump (right column).
The shapes of these optical bumps help divide WR stars into three main classes
\citep[see][and references therein]{crowther07}:
(1) WN stars: have H absorption lines and strong He and N emissions lines, 
especially in the blue bump, but can also have a few C lines; 
(2) WC stars: have strong He, O, and C emission lines, especially the \ion{C}{4} lines 
characteristic of the red bump, but lack N lines; and 
(3) WO stars: have strong lines from He and high-ionization lines from C and O. 

Using the FUV and optical spectra presented here, we can characterize the 
WR nature of the stellar populations in our sample.
We see that Regions 1 and 2 have clear \ion{N}{2} \W4621 and \ion{N}{3} \W\W4634,4641 emission in 
their blue bumps, along with weak detections of the \ion{N}{4} \W4057 feature.
Weak N features may also be present in Regions 6, 7, and 8.
We also look for features indicative of WC stars.
Just redward of the N emission in the blue bump (and blueward of [\ion{Fe}{3}]), 
a second less prominent emission feature is seen in Regions 1 and 2, but it is difficult to 
determine whether this is due to \ion{C}{3} or \ion{O}{2} emission, or both.
Further, the red \ion{C}{4} bump doesn't appear strongly in any of the \ion{H}{2} 
regions in our sample, suggesting little to no contributions from WC stars in our spectra. 
Therefore, in the five regions with strong or weak optical WR features,
we only significantly detect the blue WR bump, suggesting that WN stars are likely present.
Using the \citet{smith96} \ion{N}{5} \W4057/\ion{N}{3} 4603-4641 classification scheme,
we find these stars are most consistent with late WN6-WN8 stars in Regions 1 and 2. 

On the other hand, \citet{martins23} recently examined optical spectra of six of the 
\ion{H}{2} regions presented here for signatures of 
VMS or WR features, but did find contributions from WC stars.
However, \citet{martins23} used optical spectra from the SDSS and from GTC/OSIRIS 
from \citet{esteban20}, which have slightly different aperture pointings, sizes, and 
position angles than our CHAOS optical spectra.
Because these \ion{H}{2} regions have experienced intense star formation, they may 
contain a mixture of massive star types that are spatially distributed and so not 
equally captured in the different apertures used. 
In the present study, we took special care to design our {\it HST}/COS UV 
observations to have the same precise coordinates as the CHAOS optical spectra by 
employing target offsets from nearby point-sources.
Therefore, the CHAOS optical spectra represent the closest spatial map to the 
FUV {\it HST}/COS spectra.
However, there may still be differences in the stellar light captured between 
the FUV and optical spectra owing to the different aperture shapes and the 
vignetting of the COS aperture.
These differences in observational setup may also be the source of discrepancies 
in identifying WR and VMS features.
For example, \citet[][and, subsequently, \cite{martins23}]{esteban20} observe both 
the blue and red WR features in Region 7 (NGC 5455) and suggest that both WN and WC 
stars are present, while our spectra show no definitive signatures of WC stars. 

\subsubsection{FUV Characteristics}
\begin{figure}
\begin{center}
	\includegraphics[width=0.45\textwidth,trim=-2mm 0mm 0mm 0mm,clip]{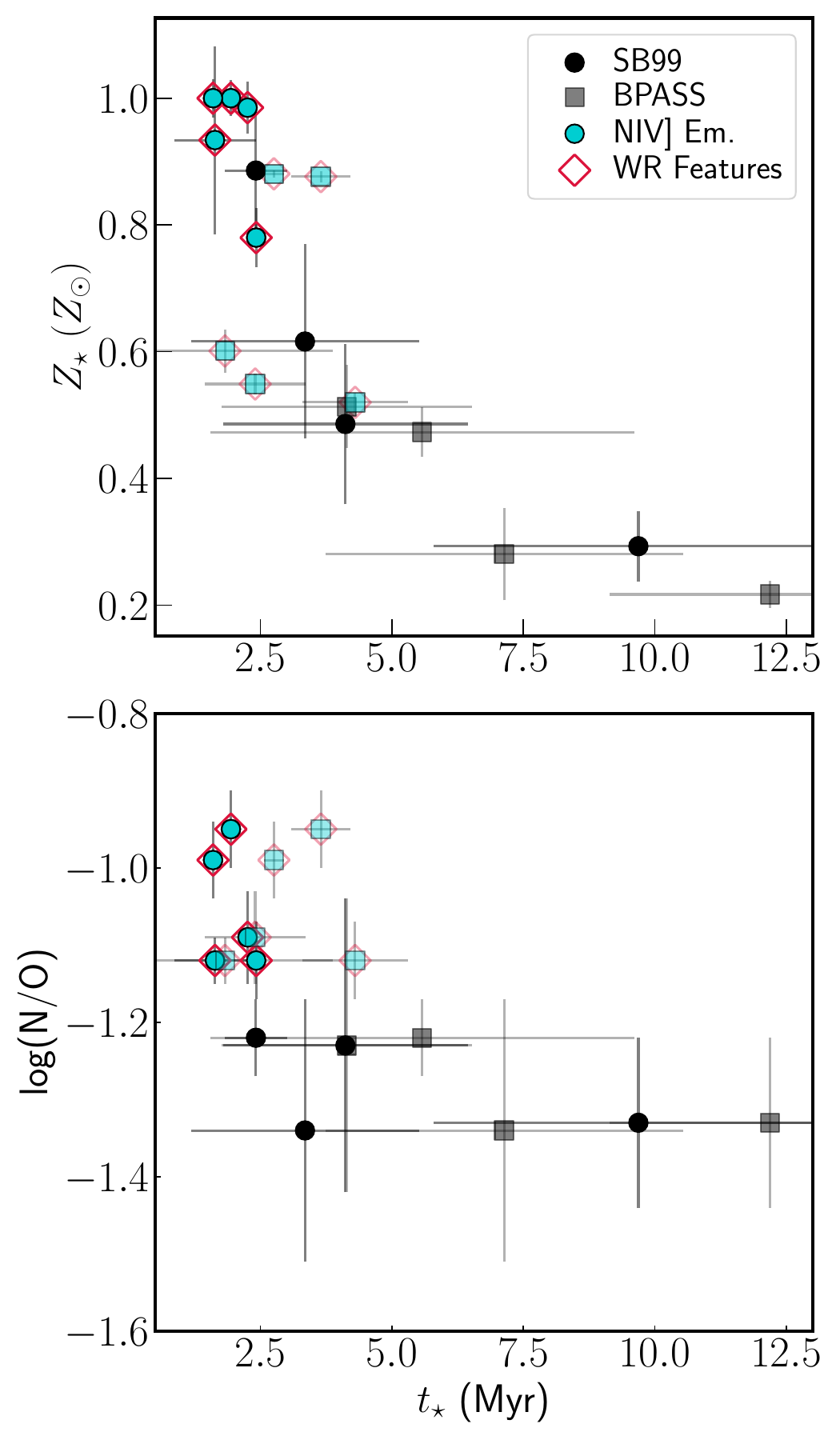} 
\caption{{\it Top:} Modeled luminosity-weighted metallicity versus age  
for the ionizing stellar populations in our \ion{H}{2} region sample.
We show points derived using both \texttt{SB99} (circles) and \texttt{BPASS}
(semi-transparent squares) models.
Regions with WR features in their optical spectra are outlined by a red
diamond; these regions also have \ion{N}{4} emission in their FUV spectra
(turquoise points).
\ion{H}{2} regions with WR features only occupy the upper left young-age,
high-metallicity regime of the plot.
{\it Bottom:} N/O abundance measured from the optical spectra versus modeled stellar age 
shows that the same regions that have enhanced N/O gas abundances have enhanced \ion{N}{4}] 
emission and strong WR features in their spectra.
This supports the idea that the WR phase is responsible for the observed local N 
enhancement in these \ion{H}{2} regions.
\label{fig9}}
\end{center}
\end{figure}

In agreement with past studies, the prominent WR emission in the M101 optical spectra 
corresponds to FUV spectra of young, metal-rich stellar populations (see Table~\ref{tbl2}).
Interestingly, the M101 observations presented here suggest a threshold for strong WR 
emission of stellar $Z \gtrsim 0.78\ Z_\odot$ ($Z\gtrsim 0.54\ Z_\odot$) and 
$t \lesssim 3$ Myrs ($t \lesssim 5$ Myrs) for the \texttt{SB99} (\texttt{BPASS}) 
models; slightly younger and less metal-rich than previously thought 
\citep[e.g.,][]{groh14}.
This is shown in Figure~\ref{fig9} where we compare the stellar metallicities and 
ages of the full \ion{H}{2} region sample to that of the WR subsample.
\ion{H}{2} regions with WR features only occupy the upper left young-age,
high-metallicity regime of the plot.

Such young ages for WR stars, as indicated by the \texttt{SB99} models, are somewhat 
surprising given that the main sequence H-core burning phase of massive stars that 
precedes the WR phase typically lasts longer than 3 Myr 
\citep[e.g., 3.5 Mr for a non-rotating 60 M$_\odot$ star][]{ekstrom12}.
However, the situation is further complicated by the fact that massive star and
WR evolutionary tracks are metallicity-dependent.
For example, \citet{leitherer95} used evolutionary synthesis models for populations of 
massive stars to show that WR stars can appear after just 2.5 Myr in twice-solar populations
and after just 2.0 Myr in solar populations.
\citet{eldridge17} also examined the presence of WR stars by subclass and found that
WC stars appear after 2.5 Myr in solar populations, in agreement with \citet{leitherer95},
but that WN stars can appear slightly earlier after 2.0 Myr.
If WR stars can appear after only 2 Myr, then, within the uncertainties, the stellar 
population ages for our sample allow for the presence of WRs.
Only the \texttt{SB99} age for Region 2 would be exceptional, but is also within agreement
assuming an accuracy of $<1$ Myr \citep{sirressi22}. 

Alternatively, we could be observing a pre-WR phase of VMSs.
The characteristic emission features used to identify WR stars arise in their dense, 
high-velocity stellar winds.
Because VMSs also produce exceptionally strong stellar winds and have high mass loss rates,
their spectra can mimic those of the later evolutionary stage of WRs 
\citep[e.g.,][]{martins08,grafener08}.
In general, the FUV spectra of VMSs and WRs are difficult to tell apart.
Both VMSs and WRs have been observed to have  many of the features present in our
{\it HST}/COS spectra, including \ion{N}{4}] \W1486, \ion{He}{2} \W1640, and 
\ion{N}{4} \W1720 \citep[e.g.,][]{martins22}.
On one hand, \citet{martins23} argue that the UV alone is insufficient and the optical WR 
feature morphologies are needed to distinguish between VMSs and WRs.
As such, the presence of the blue bump in some of our spectra favors the presence of WR stars.
On the other hand, VMSs are only present at very young stellar population ages 
($<2.5$ Myr), suggesting VMSs could be the source of the excess wind profiles in our 
spectra and the corresponding young ages of the \texttt{SB99} fits.

The ionizing stellar populations in M101 could contain a complicated mixture of massive 
stars with a range of ages ($<10$ Myr) such that both VMSs and WRs are present.
However, the presence of VMSs alone is deemed unlikely.
First, the presence of \ion{Si}{4} P-Cygni profiles in Regions 1 and 2 in Figures~\ref{fig2} 
and \ref{fig3} require older ages than what is physical for VMSs 
\citep[e.g.,][]{crowther16,smith16}.
Additionally, the presence of \ion{O}{5} \W1371 is considered a key spectral diagnostic 
of VMSs, but there are no strong detections in the M101 spectra.
The observed spectrum of Region 1 in Figure~\ref{fig3} shows weak \ion{O}{5} absorption,
indicating that VMSs may also be present in this region.
However, the other regions show either no absorption at \ion{O}{5} \W1371 or
blue-shifted absorption that is more consistent with \ion{Fe}{5} \W1365.
For the other WR spectra (Regions 2, 6, 7, and 8) in Figure~\ref{fig3}, the ambiguous 
absorption feature near \W1370 is well fit by the \ion{Fe}{5} feature in the stellar 
continuum model.
Therefore, WRs seem to be the dominant wind source in our spectra, although we cannot 
Further, such large, complex \ion{H}{2} regions as are seen in M101 might be
likely places to find multiple, unresolved stellar populations hosting both WRs and VMS.

Proving the presence of strong WR stars at moderate metallicities is also challenging
because WR features have not been observed in the spectra of individual metal-poor
massive stars. 
This may simply be due to the fact that metal-driven winds in the photospheres of 
massive stars decrease with lower metallicities by definition. 
As a result, WR features have also proven difficult to produce in theoretical models 
at low  metallicities, while WR features appear at relatively high metallicities and 
progressively increase in strength (e.g., broad \ion{He}{2} emission) in the stellar 
population synthesis models.
Therefore, the presence of WR features in our observed spectra may be driving our 
best continuum fits to higher luminosity-weighted metallicities.
In turn, requiring high metallicities to fit the WR features may drive the 
luminosity-weighted ages younger.
This could explain the large offset in \texttt{BPASS} stellar metallicity in 
Figure~\ref{fig7} for the subset of \ion{H}{2} regions with WR features in their spectra.

The WN-type stars in our M101 \ion{H}{2} region sample are further confirmed by their
relatively large N/O abundances in Table~\ref{tbl2} (larger for WR regions than 
non-WR regions) and their FUV N spectral features.
As mentioned in Section~\ref{sec:contfit}, WR stars may also be responsible for the 
\ion{N}{4}] \W\W1483,1487 emission and excess emission in the \ion{N}{4} \W1719 P-Cygni 
profile are seen in Figures~\ref{fig2} (also see Figure~\ref{fig4}).
Such \ion{N}{4}] detections are very intriguing.
First, \ion{N}{4}] \W\W1483,1487 emission is rare and has only been reported 
for a handful of galaxies and active galactic nuclei (AGN) across all redshifts.
Second, there has been heightened interest in \ion{N}{4}] due to its strong detection
in the {\it JWST}/NIRSpec spectrum of the exceptionally bright $z=10.6$ galaxy 
GN-z11 \citep{bunker23}.
As a result, WR stars have been suggested as a source of increased N emission for
GN-z11 and other galaxies \citep[e.g.,][]{senchyna24}.

The WR-driven N features seen in our M101 spectra are further supported by the presence 
of P-Cygni profiles of the resonance or excited transitions of metal lines in bluest 
portions of the FUV spectra, including \ion{S}{6} \W\W933,945, \ion{N}{4} \W955, 
\ion{N}{3} \W991, \ion{O}{6} \W\W1032,1038, \ion{P}{5} \W\W1118,1128, and \ion{C}{3} \W1175.
This combination of features and range of moderate- to high-ionization potential species 
is consistent with the later WN6-WN7 {\it FUSE} spectra described in \citet{willis04}.
Additionally, if WC stars were present, we'd expect to see strong emission or P-Cygni
profiles of \ion{C}{3} \W\W977,1140 and/or \ion{C}{4} \W\W1107,1135,1169, yet we 
see no evidence of significant WC stars in the blue segment of our FUV spectra. 

\subsection{Impact of WR Stars on Inferred Stellar Properties}
While our integrated FUV spectra are difficult to fit with current stellar models 
owing to the complex blend of massive star and WR contributions, it is clear
that the WR features can have a significant impact on stellar continuum and
subsequently derived properties.
The impact of WR stars on integrated star-forming spectra can be seen in the 
residuals to our stellar population model fits.
The spectral regions where our stellar population synthesis fits perform the 
poorest correspond to underfit WN P-Cygni and emission profiles (e.g., \ion{N}{4}~\W1720), 
suggesting that the
\texttt{SB99} and \texttt{BPASS} models fail to reproduce the full WR
populations in these observed regions.
Fortunately, the Potsdam Wolf-Rayet \citep[\texttt{PoWR};][]{sander15,todt15} 
model atmospheres for WR stars can provide significantly better fits to the 
FUV WN features.
In the yellow window of Figure~\ref{fig3} we expand the observed spectrum of Region 1 
around the \ion{N}{3} \W991 feature and overplot a stellar model from the \texttt{PoWR} 
MW WNL-H20 grid that corresponds to $T_\star = 50.1$ kK and log $(R_t/R_\odot) = 0.3$.
After a relative scaling between spectra, it is clear that the \ion{N}{3} \W991 
emission is easily reproduced by a WN star.
Therefore, the observed spectra require a combination of ordinary massive stars 
and WN stars to full fit it.

On the other hand, the WN features discussed in this work can provide
powerful tools to improve our interpretation of FUV continuum spectra.
Using our M101 FUV spectra, we have demonstrated that the $\sim900-1200$ \AA\ 
stellar continuum features provide strong diagnostics of the presence of WR stars 
that can be used to interpret the integrated spectra of star-forming regions or galaxies. 
Further, these very blue spectra provide support for the presence of WN stars 
as the source of excess \ion{N}{4} \W\W1483,1486 and \ion{N}{4} \W1720 emission.
As discussed above, the presence of very high ionization stellar wind profiles 
suggests the presence of very-hot / young massive stars, whose integrated spectra are 
better characterized by the \texttt{SB99} and \texttt{BPASS} models.
However, the \texttt{SB99} and \texttt{BPASS} models either do not have sufficient numbers of WR stars
or do not include appropriate WR evolutionary tracks to reproduce the excess
N emission observed in our spectra.
As a result, in trying to fit this excess emission,
the \texttt{SB99} model fits are biased to higher-metallicities and younger ages.
Therefore, the \ion{H}{2} regions of M101 provide excellent benchmarks for 
constraining the revised inclusion of WR and VMS stars in stellar population models.
As such, we provide the emission-line properties for important UV and optical WR 
features measured for Regions 1 and 2 in Table~\ref{tbl4}.


\begin{deluxetable}{rcc c cc}
\tabletypesize{\small}
\tablecaption{Stellar Wind Emission-Line Properties}
\tablehead{
\CH{}       & \multicolumn{2}{c}{Region 1} && \multicolumn{2}{c}{Region 2} \\ \cline{2-3} \cline{5-6} 
\CH{}       & \CH{EW}    & \CH{FWHM}          && \CH{EW}    & \CH{FWHM} \\ [-1.5ex]
\CH{Line}   & \CH{(\AA)} & \CH{(km s$^{-1}$)} && \CH{(\AA)} & \CH{(km s$^{-1}$)}}
\startdata
\ion{N}{4}] \W1486     & $-1.37$ & 1434 && $-0.72$ & 862  \\
\ion{He}{2} \W1640     & $-2.77$ & 1436 && $-3.08$ & 1344 \\
\ion{N}{4} \W1720      & $-1.08$ & 603  && $-1.54$ & 636  \\
? \W1726               & $-0.77$ & 603  && $-0.81$ & 636  \\
\ion{N}{3} \W4634      & $-0.99$ & 950  && $-1.78$ & 936  \\
\ion{N}{3} \W4640      & $-2.00$ & 1085 && $-2.36$ & 748  \\
\ion{C}{3} \W4650      & $-0.33$ & 345  && $-0.57$ & 294  \\
\ion{He}{2} \W4686     & $-3.20$ & 975  && $-8.99$ & 1586 \\
\ion{C}{4} \W5801      & $-0.55$ & 775  && $-0.27$ & 362  \\
\ion{C}{4} \W5808      & $-0.55$ & 775  && $-0.38$ & 362  \\ [0.5ex]
\hline\hline 
\\ [-1.0ex]
{}       & \multicolumn{2}{c}{\ion{He}{2} \W1640} 
        && \multicolumn{2}{c}{\ion{He}{2} \W4686} \\ \cline{2-3} \cline{5-6} 
{}       & EW      & FWHM && EW      & FWHM \\ 
{}       & (\AA)   & (km s$^{-1}$) && (\AA) & (km s$^{-1}$) \\ \hline
Region 1 & $-2.77$ & 1436 && $-3.20$ & 975  \\
Region 2 & $-3.08$ & 1344 && $-8.99$ & 1586 \\
Region 4 & --      & --   && --      & --   \\
Region 5 & $-1.81$ & 742  && --      & --   \\
Region 6 & $-2.05$ & 1113 && $-3.73$ & 1202 \\
Region 7 & $-2.62$ & 1981 && $-3.52$ & 1574 \\
Region 8 & $-3.34$ & 1666 && $-3.36$ & 1201 \\
Region 9 & $-1.02$ & 859  && $-0.68$ & 289  \\
Region 10 &$-1.95$ & 1420 && $-1.31$ & 296  \\
\enddata	
\tablecomments{ 
Emission-line properties of stellar wind features observed in the UV and optical 
spectra of the M101 sample.
{\it Top:}
Line fits for the regions with the strongest wind features (Regions 1 and 2). 
Column 1 lists the ion and wavelength of each emission line, with the exception of
the emission feature near \W1726 that has an uncertain association.
Columns 2 and 3 list the equivalent width (EW) and Gaussian full width half maximum 
(FWHM) for Region 1, while Columns 4 and 5 list EW and FWHM for Region 2.
{\it Bottom:}
UV and optical \ion{He}{2} emission properties for the full sample.
Note that the \ion{He}{2} features are not well fit in the low-resolution UV spectra
for regions with weak emission (EW$_{\lambda1640} < 2$\AA). 
}
\label{tbl4}
\end{deluxetable}


\begin{figure}
\begin{center}
	\includegraphics[width=0.5\textwidth,trim=0mm 20mm 0mm 45mm,clip]{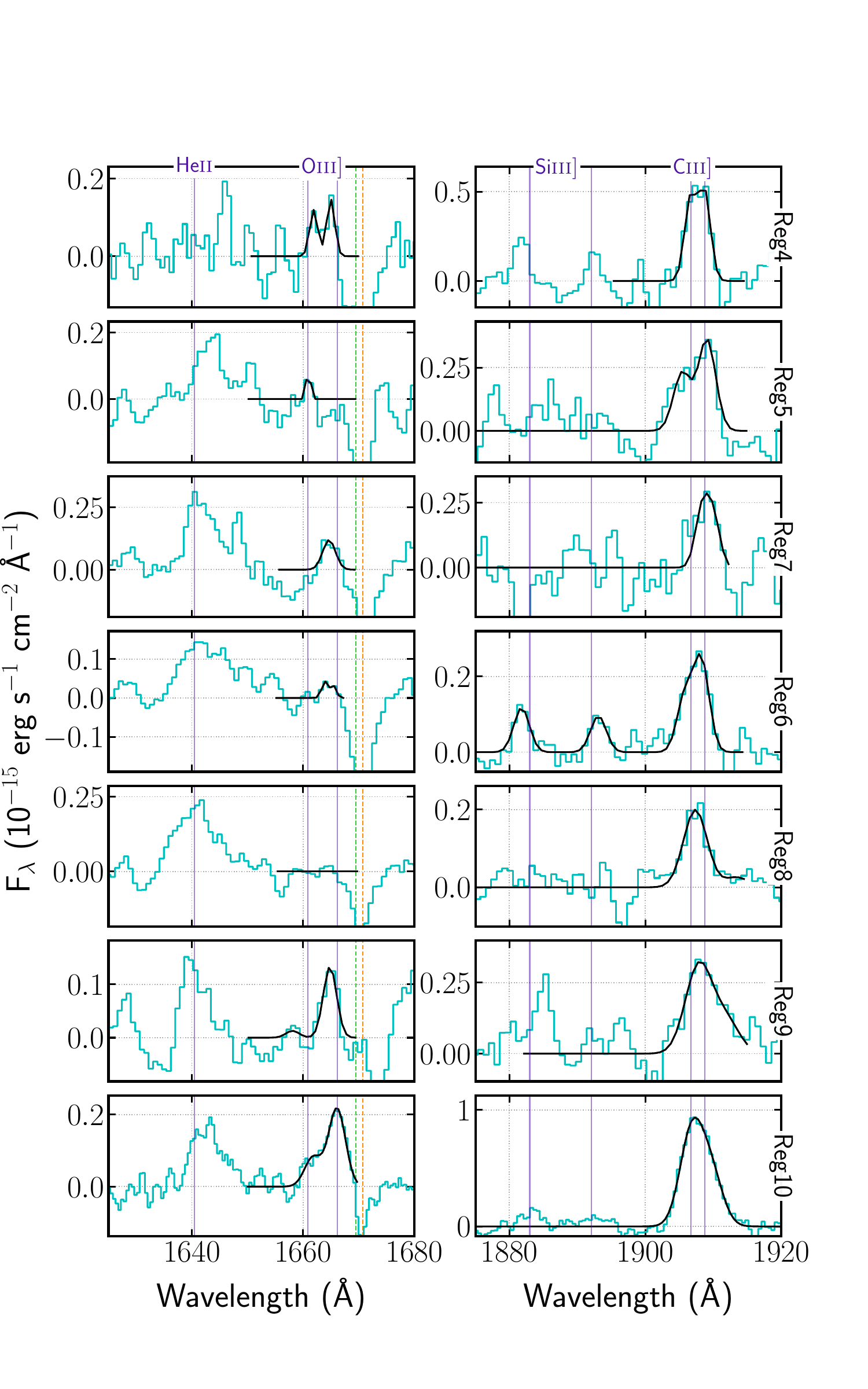}
\caption{ 
Continuum-subtracted {\it HST}/COS spectra of the seven 
\ion{H}{2} regions with \ion{C}{3}] \W\W1907,1909 emission.
Monte-Carlo minimization fits to the Gaussian profiles of \ion{O}{3}] and
\ion{C}{3}] are shown as black lines.
Many of the \ion{O}{3}] profiles lack sufficient signal due to the neighboring 
\ion{Al}{2} \W1671 ISM (orange dashed line) and MW (green dashed line) 
absorption profiles. \label{fig10}}
\end{center}
\end{figure}


\subsection{\ion{O}{3}] and \ion{C}{3}] Emission}\label{sec:co}
The FUV \ion{O}{3}] \W\W1661,1666 and \ion{C}{3}] \W\W1907,1909 nebular emission lines
offer the opportunity to investigate the relative C/O abundance in galaxies. 
In general, carbon abundance determinations in extragalactic \ion{H}{2} regions are rare 
in the literature due to the lack of bright CELs of C 
in the optical and the extreme faintness of the optical RLs. 
Previous C/O studies have been most successful in measuring the UV C and O CELs in 
metal-poor dwarf galaxies \citep[12+log(O/H) $<$ 8.0); e.g.,][]{garnett95,berg16,berg19b}, 
while optical RL studies dominate in metal-rich regions 
(12+log(O/H) $>$ 8.0) of very nearby galaxies 
\citep[e.g.,][]{peimbert05,bresolin07,esteban09,esteban14,toribiosancipriano16,toribiosancipriano17}. 
Given this apparent dichotomy in the previous C/O measurements, comparisons between the 
CEL and RL methods for C/O have remained sparse. 

Optical 
measurements of abundances in metal-rich \ion{H}{2} regions (12+log(O/H) $>$ 8.0) typical 
of spiral galaxies are difficult in general. 
Since the cooling in \ion{H}{2} regions is dominated by IR fine-structure lines of N$^{++}$ 
and O$^{++}$ for high metallicities, the greater the metallicity, the greater the cooling. 
The lower electron temperatures mean that the higher excitation CELs become progressively 
weaker, and the electron temperature becomes difficult to measure. 
This is a problem especially for C, as the main ionization state, C$^{++}$, has its 
strongest emission in the UV, and this becomes vanishingly faint in metal-rich nebulae. 

Due to these challenges, there is a paucity of metal-rich targets with 
significant UV \ion{C}{3}] and \ion{O}{3}] detections.
For example, \citet{garnett99} observed \ion{C}{3}] in 4 metal-rich \ion{H}{2} regions 
in the spiral galaxies NGC 2403 and M101 using FOS/{\it HST}, but were unable to obtain any 
significant detections of the \ion{O}{3}] doublet. 
Because COS is significantly more sensitive than FOS, {\it HST} is now better able to detect 
the \ion{C}{3}] and \ion{O}{3}] UV emission lines in metal-rich \ion{H}{2} regions,
such as the two regions with 12+log(O/H) $>$ 8.2 reported in \citet{senchyna17}. \looseness=-2

\subsubsection{Flux Measurements} \label{sec:lines}
The \ion{O}{3}] \W\W1661,1666 and \ion{C}{3}] \W\W1907,1909 emission line features of our 
UV {\it HST}/COS spectra are plotted in Figure~\ref{fig10}.
To determine the emission line strengths for these UV features, we first continuum-subtracted 
the spectra using the best-fit stellar continua described in Section~\ref{sec:continuum}.
The UV emission lines were then fit using the \texttt{scipy.optimize.curve\_fit} \citep{virtanen20}
to perform a least-squares minimization on the built-in Gaussian function in 
the \texttt{lmfit} \texttt{python} package \citep{newville18}. 
Groups of nearby lines were fit simultaneously, constrained by a single Gaussian 
FWHM and a single line center offset from the vacuum wavelengths (i.e., redshift).
Figure~\ref{fig10} shows the Gaussian fits to the significant \ion{O}{3}] and \ion{C}{3}] 
emission lines in the COS spectra of M101.

The errors of the flux measurements were approximated using
\begin{equation}
	\sigma_{\lambda} \approx \sqrt{(2 \times \sqrt{N} \times \mbox{RMS})^2 + (0.01 \times F_{\lambda})^2},
	\label{eq1}
\end{equation}
where N is the number of pixels spanning the Gaussian profile fit to the emission lines. 
The root mean squared (RMS) noise in the continuum was taken to be the average of the RMS on 
each side of an emission line when free of absorption features. 
The two terms in Equation~\ref{eq1} approximate the errors from continuum subtraction and flux calibration,
where the flux calibration uncertainty is assumed to be 1\%\ of the flux based on 
the standard star calibrations in CALSPEC \citep{bohlin10}.
For weak lines, such as the UV CELs, the RMS term determines the approximate uncertainty. 
Finally, UV fluxes and errors were corrected for extinction using the reddening law of 
\citet{reddy16} and nebular color excesses ($E(B-V)$) derived from the optical Balmer lines 
from \citet{croxall16}.
The reddening corrected line intensities measured for our C/O targets are reported in 
the top of Table~\ref{tbl5}. 

Seven of our 10 targets have significant \ion{C}{3}] detections, 
which we consider a high success rate given the relatively large O/H abundances of our sample.
However, only one region has a significant \ion{O}{3}] detection.
For nearby galaxies at redshifts close to 0.003, it is possible for the \ion{O}{3}] \W1666 
to be significantly affected by absorption from the Galactic \ion{Al}{2} \W1671 line 
\citep[see, e.g., the UV spectrum of SB191 in][]{senchyna17}. 
M101 has a redshift of $z\approx0.0008$ such that the \ion{O}{3}] \W1666.15 emission line 
shifts to a line center of 1667.48 \AA, but \ion{O}{3}] can be further shifted towards 
\ion{Al}{2} contamination by the rotational velocity of a given \ion{H}{2} region within M101.
These effects, combined with the low resolution of our spectra, cause the \ion{Al}{2} 
absorption to significantly affect the \ion{O}{3}] profile.
The resulting effect is clearly seen in Figure~\ref{fig10}:
the \ion{O}{3}] line strengths are weak for most of the sample, where the blue \ion{Al}{2}
absorption wing is clearly blended with the \ion{O}{3}] emission. 
Therefore, higher resolution spectra are needed to strongly detect the FUV \ion{O}{3}]
nebular emission in M101 and other metal-rich targets. 


\subsubsection{The C/O Abundance Ratio}\label{sec:UV_CO}
Determining relative abundances requires significant ($>3\sigma$) line detections.
While several of our \ion{C}{3}] line detections meet this criteria, only Region 10 has
a significant \ion{O}{3}] detection.
We, therefore, perform C/O abundance determinations only as an exploratory exercise,
noting that higher S/N and higher resolution spectra are needed for robust measurements.
Given the faint hints of \ion{O}{3}] emission in Regions 9, 7, 6, and 4, significant
detections could likely be achieved with higher spectral resolution.
We note that we refrain from performing a 1-sigma upper limit 
calculation because the contamination of the MW \ion{Al}{2} \W1671 absorption feature 
would lead to an underestimation of the upper limit.\looseness=-2

To perform an exploratory C/O abundance analysis, 
an initial approximation of C/O can be determined from 
the C$^{+2}$/O$^{+2}$ ratios reported in Table~\ref{tbl5}. 
However, the higher ionization potential of O$^{+2}$ relative to C$^{+2}$ (54.9 eV 
versus 47.9 eV, respectively), means that an ionization correction factor is needed 
to account for any unseen ionization states.
For the spectra shown in Figure~\ref{fig2}, the \ion{C}{4} emission features all seem 
to be well fit by a stellar wind P-Cygni profile, suggesting that we are not missing any 
significant emission from C$^{+3}$ ions. 
However, there could be contributions from lower ionization states. 
Therefore, we determine C/O abundances using:
\begin{minipage}{0.47\textwidth}
\begin{align}
	{\frac{\mbox{C}}{\mbox{O} }} & = {\frac{\mbox{C}^{+2}}{\mbox{O}^{+2}}\ }\times \Bigg[{\frac{X(\mbox{C}^{+2})}{X(\mbox{O}^{+2})}}\Bigg]^{-1} \nonumber \\
			     			    & = {\frac{\mbox{C}^{+2}}{\mbox{O}^{+2}}\ }\times{\mbox{C ICF}},	
\end{align}
where X(C$^{+2}$) and X(O$^{+2}$) are the C$^{+2}$ and O$^{+2}$ volume fractions, respectively.
Following the metallicity-dependent ICF method of \citet{berg19b}, 
we ran {\sc cloudy} 17.00 \citep{ferland13} photoionization models to match the properties 
spanned by M101. 
We used \texttt{BPASS} burst models for the input ionizing radiation field, 
a range of burst ages ($10^{6.0}-10^{7.0}$ yrs), 
ionization parameter ($-3.0 <$ log $U <-1.0$), and
metallicities ($Z = 0.004-0.02 = 0.2-1.0\ Z_{\odot}$),
with matched stellar and nebular metallicities ($Z_{\star} = Z_{neb}$).
\end{minipage}

\begin{figure}
\begin{center}
 \includegraphics[width=0.4\textwidth,trim=0mm 0 0 0,clip]{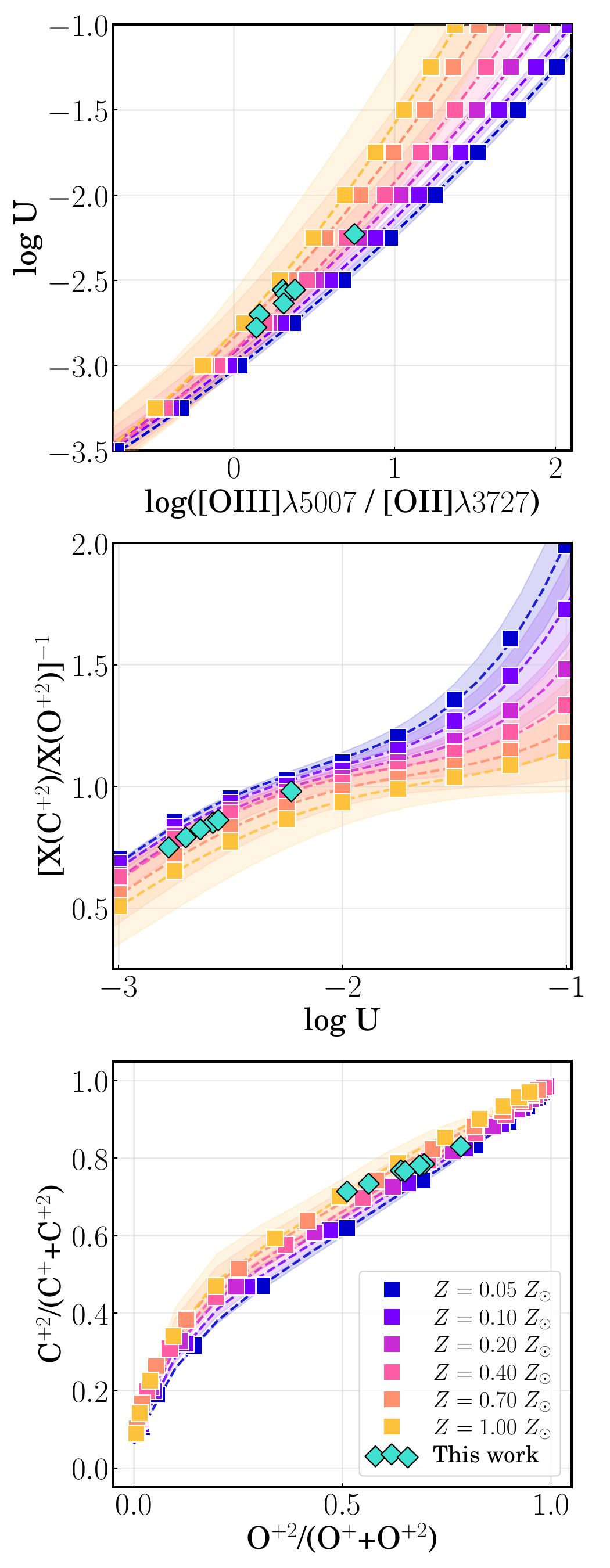} 
\caption{
Photoionization models used to derive the carbon ICF.
Models with $t=10^{6.5}$ yrs are plotted as squares and color-coded by their metallicity. 
The best-fit polynomial function for each metallicity is plotted as a dashed line;
fit coefficients are reported in Table~\ref{tbl6}.
The range of models spanned by ages of $t=10^{6-7}$ yrs are designated 
by the shaded bands, representing an estimate of the uncertainty.
{\it Top:} Ionization parameter, log $U$, as a function of the 
[\ion{O}{3}] \W5007/[\ion{O}{2}] \W3727 emission line ratio. 
{\it Middle:} Carbon ionization correction factor, 
CICF = [X(C$^{+2}$)/X(O$^{+2}$)]$^{-1}$, as a function of log $U$.
{\it Bottom:} Another CICF, C$^{+2}$/(C$^+$+C$^{+2}$),
but derived as a function of the observed O$^{+2}$/(O$^+$+O$^{+2}$) ion fraction.
The observed line ratios and ion fractions from \citet{croxall16} and derived 
log $U$ and ICF values are plotted as teal diamonds. 
\label{fig11}}
\end{center}
\end{figure}

Our {\sc cloudy} models are shown in Figure~\ref{fig11} comparing the [\ion{O}{3}] 
\W5007/[\ion{O}{2}] \W3727 emission line ratios, ionization parameters, and 
ionization fractions of C$^{+2}$ and O$^{+2}$.
The top panel shows the increasing trend of [\ion{O}{3}] \W5007/[\ion{O}{2}] 
\W3727 with ionization parameter, which is a useful diagnostic. 
The points are color-coded by model stellar/nebular metallicity, while the 
shaded band denotes the corresponding models spanned by the range in burst age.
We fit each of the metallicity models with a polynomial of the shape:
log $U = c_3\cdot{x^2} + c_2\cdot{x} + c_1$, where 
$x =$ log([\ion{O}{3}] \W5007/[\ion{O}{2}] \W3727).
The coefficients for these fits are listed in Table~\ref{tbl6}.
We note that the coefficients derived in this work are slightly different than 
those reported in \citet{berg19b} for common metallicities because we took the 
average values of the models spanning different ages, while the previous work 
fit the $t=10^{6.5}$ yr models only.
Using the observed [\ion{O}{3}] \W5007/[\ion{O}{2}] \W3727 ratios from 
\citet{croxall16} and the metallicities reported in \citet{berg20}, we determined 
log $U$ values by interpolating between the polynomial fits defined for each 
metallicity steps; the resulting values are reported in Table~\ref{tbl5}.

With log $U$ estimates in hand, we determined C ICFs for our sample. 
As shown in the middle panel of Figure~\ref{fig11}, both metallicity and 
stellar age have moderate effects on the relative ionization fraction of C.
We fit each of the metallicity models with a polynomial of the shape:
C ICF = [X(C$^{+2}$)/X(O$^{+2}$)]$^{-1}$ = 
$c_5\cdot{x^4} + c_4\cdot{x^3} + c_3\cdot{x^2} + c_2\cdot{x} + c_1$, 
where $x =$ log $U$.
The coefficients for these fits are listed in Table~\ref{tbl6}.
We determined moderate C ICFs for our sample, showing a moderate 
to small correction for our sample that lies between log$U=-2.75$ to $-2.30$.
We estimated the uncertainty in the ICF as the scatter amongst the different
models considered (relative abundances and burst age) at a given log$U$. 
Ionization parameters, C ICFs, ionic C fractions, and corrected C/O ratios 
are provided in Table~\ref{tbl5}. 


\begin{deluxetable*}{lCCCCCCCCC}
\tabletypesize{\scriptsize}
\tablecaption{C/O Emission-Line and Abundance Estimates}
\tablehead{
\CH{Property}       & \CH{Reg 4}    & \CH{Reg 5}    & \CH{Reg 6}    & \CH{Reg 7}    & \CH{Reg 8}    & \CH{Reg 9}    & \CH{Reg 10}}
\startdata	   
$I($\ion{O}{3}])/$I($\ion{C}{3}])
                    & 0.144\pm0.129 & 0.078\pm0.253 & 0.163\pm0.162 & 0.073\pm0.177 & 0.192\pm0.328 & 0.198\pm0.233 & 0.183\pm0.014 \\
E(B--V)             & 0.185         & 0.093         & 0.042         & 0.180         & 0.070         & 0.049         & 0.231        \\
\hline 
$T_e$[\ion{O}{3}] (K)& 9540\pm90    & 9180\pm90     & 9440\pm70     & 9300\pm70     & 9580\pm80     & 10690\pm90    & 12790\pm170  \\
C$^{+2}$/O$^{+2}$   & 0.21\pm0.68   & 0.36\pm0.50   & 0.18\pm0.24   & 0.48\pm0.44   & 0.16\pm0.44   & 0.17\pm0.28   & 0.23\pm0.10  \\
log $O_{32}$        & 0.16\pm0.02   & 0.31\pm0.01   & 0.32\pm0.01   & 0.14\pm0.01   & 0.38\pm0.01   & 0.31\pm0.01   & 0.75\pm0.02  \\
log $U$             & -2.70\pm0.09  & -2.56\pm0.10  & -2.58\pm0.09  & -2.78\pm0.08  & -2.55\pm0.08  & -2.64\pm0.06  & -2.23\pm0.06 \\
C ICF               & 0.791         & 0.862         & 0.851         & 0.750         & 0.862         & 0.824         & 0.981        \\
C$^{+2}$/C ICF      & 0.714         & 0.768         & 0.785         & 0.734         & 0.781         & 0.766         & 0.830        \\
log(C/O)&{\it -0.79\pm0.63}&{\it -0.51\pm0.87}&{\it -0.81\pm0.36}&{\it -0.45\pm0.67}&{\it -0.87\pm0.58}&{\it -0.85\pm0.43}& -0.65\pm0.16\\
log(N/O)            & -1.23\pm0.19  & -1.22\pm0.05  & -1.12\pm0.03  & -1.12\pm0.05  & -1.09\pm0.06  & -1.34\pm0.17  & -1.33\pm0.11 \\
12+log(O/H)         & 8.51\pm0.10   & 8.51\pm0.02   & 8.45\pm0.02   & 8.39\pm0.01   & 8.36\pm0.02   & 8.29\pm0.06 & 8.16\pm0.02
\enddata	
\tablecomments{ 
The top three rows give the reddening-corrected intensities of \ion{O}{3}] 
and \ion{C}{3}] from the UV spectra relative to H$\beta$ from the optical 
spectra, determined using the E(B--V) reddening reported in \citet{croxall16}.
Note that the \ion{O}{3}] flux is just the strong line of the doublet, the 
\W1666 feature, but the \ion{C}{3}] flux is the total of the \W\W1907,1909 lines. 
The bottom portion of the table lists the physical properties derived for the gas,
where optical-only properties are reported from \citet{croxall16} and \citet{berg20},
and UV-derived properties are from this work.
Only Region 10 has significant detections of both \ion{O}{3}] and \ion{C}{3}];
values for the other six regions are only estimates and are italicized to 
denote this. 
\label{tbl5}}
\end{deluxetable*}


\begin{deluxetable}{crrrrrr}
\tablecaption{Coefficients for ICF Model Fits}
\tablehead{
\multicolumn{1}{c}{} 	& \multicolumn{6}{c}{$Z(Z_{\odot})$} \\ 
\cline{2-7}
\CH{$y = f(x)$} 	  & \CH{0.05} & \CH{0.10}& \CH{0.20} & \CH{0.40} & \CH{0.70} & \CH{1.00}}
\startdata	
\multicolumn{1}{l}{\bf{log $U$:}} & & & & & & \\
\multicolumn{1}{l}{$x =$ log $O_{32}$} & & & & & & \\
{$c_3$ ..............}  & 0.0798    & 0.0899    & 0.1045  & 0.1308  & 0.1592  & 0.1938 \\	
{$c_2$ ..............}  & 0.7253    & 0.7514  	& 0.7932  & 0.8588  & 0.9462  & 1.0290 \\	
{$c_1$ ..............}  & -3.026 	& -2.978	& -2.914  & -2.914  & -2.836  & -2.801 \\
\multicolumn{1}{l}{\bf{C ICF:}} & & & & & &  \\
\multicolumn{1}{l}{$x =$ log $U$} & & & & & & \\
{$c_5$ ..............}	& 0.127	    & 0.111  	& 0.098  & 0.081  & 0.070 & 0.056 \\	
{$c_4$ ..............}	& 1.373   	& 1.173	    & 1.001  & 0.815  & 0.671 & 0.528 \\	
{$c_3$ ..............}	& 5.298   	& 4.389 	& 3.558  & 2.807  & 2.190 & 1.643 \\	
{$c_2$ ..............}	& 9.070 	& 7.267	    & 5.575  & 4.246  & 3.151 & 2.286 \\
{$c_1$ ..............}	& 7.023	    & 5.681	    & 4.415  & 3.520  & 2.792 &	2.264 
\vspace{-2ex} 
\enddata
\tablecomments{
{\sc cloudy} photoionization model fits of the form 
$f(x)=c_5\cdot{x^4}c_4\cdot{x^3}+c_3\cdot{x^2}+c_2\cdot{x}+c_1$.
The model grid and polynomial fits are shown in Figure~\ref{fig11}.
The models are described in \S~\ref{sec:UV_CO}.
\label{tbl6}}
\end{deluxetable}

The bottom panel of Figure~\ref{fig11} shows the ionization fraction of 
C$^{+2}$/(C$^+$+C$^{+2}$) vs O$^{+2}$/(O$^+$+O$^{+2}$). 
If the singly- and doubly-ionized species are the dominant species of C and O,
then these ratios can be used directly as an independent ICF.
Using the O$^{+2}$/(O$^+$+O$^{+2}$) ratio derived from the optical spectra,
we predict the C$^{+2}$/(C$^+$+C$^{+2}$) ratios for our sample.
The C$^{+2}$/C ICF values are report in Table~\ref{tbl5}.
Compared to the our C ICFs derived from the ionization parameter, 
the C$^{+2}$/C ICF values are always lower and less than one, which would 
provide a more significant correction to our C/O abundances. 

Note that the C and O abundances presented here have not been corrected 
for the fraction of atoms embedded in dust.
\citet{peimbert10} have estimated that the depletion of O ranges between 
roughly 0.08$-$0.12 dex, and has a positive correlation with O/H abundance.
C is also expected to be depleted in dust, 
mainly in polycyclic aromatic hydrocarbons and graphite.
The estimates of the amount of C locked up in dust grains in the local 
interstellar medium shows a relatively large variation depending on the 
abundance determination methods applied \citep[see, e.g.,][]{jenkins14}.
However, carbon and oxygen are often found to have similar amounts of depletion 
onto dust grains, with corrections typically around 0.1 dex \citep[e.g.,][]{esteban98,jenkins09}.
Assuming the C and O depletions are similar, we take the gas-phase C/O ratio as
indicative of the total C/O abundance for this exercise.


\begin{figure}
\begin{center}
    \includegraphics[width=0.475\textwidth,trim=5mm 0mm 0mm 0mm,clip]{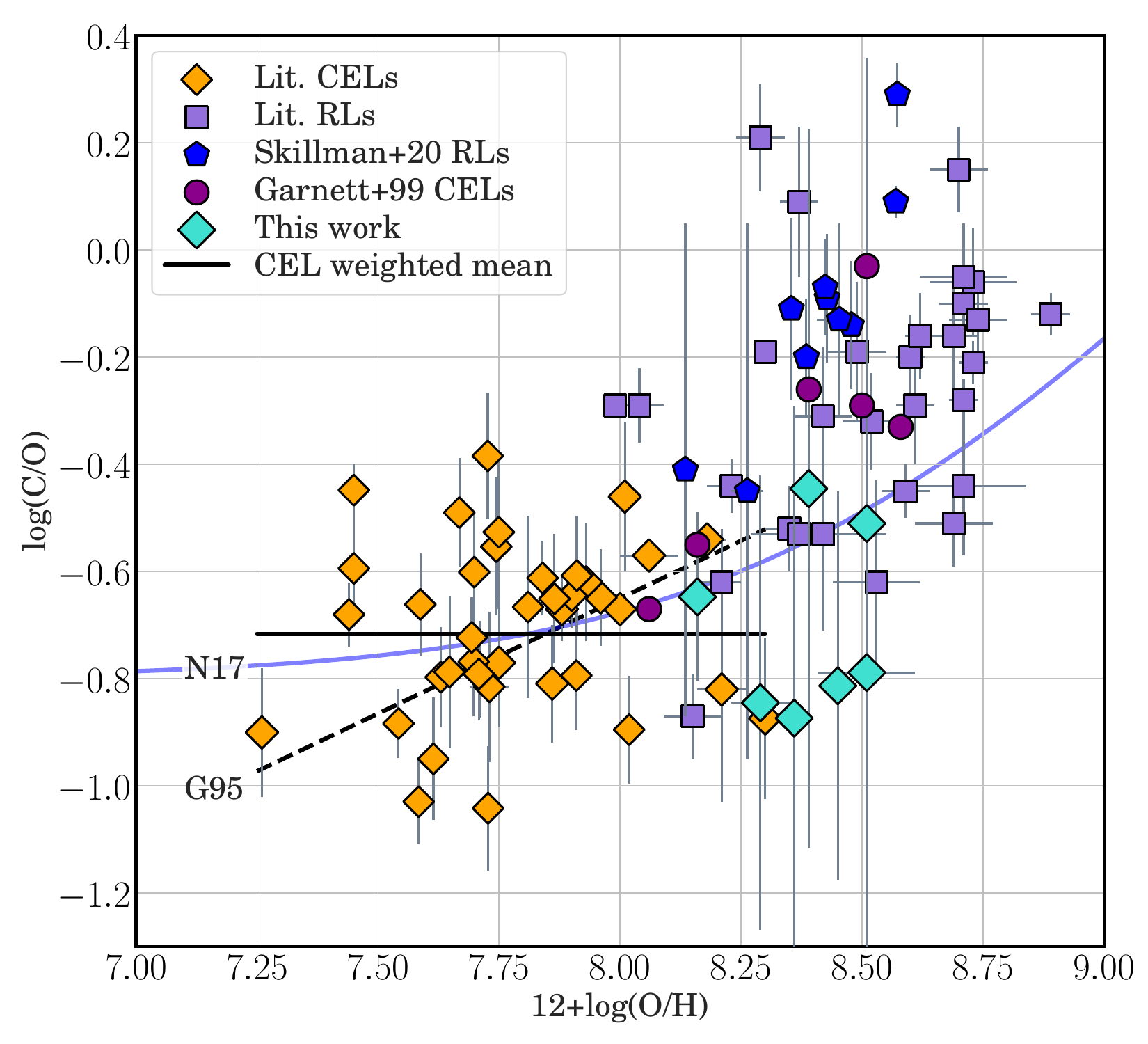} 
\caption{ 
Carbon to oxygen ratio versus oxygen abundance for star-forming galaxies
and \ion{H}{2} regions. 
Low-metallicity galaxies with CEL C/O measurements from the literature 
\citep[][]{berg16,senchyna17,berg19a} are plotted as gold diamonds.
Some additional CEL C/O abundances are plotted for individual \ion{H}{2} regions
from \citet{garnett99}.
We also include RL C/O abundances measured for higher-metallicity \ion{H}{2} regions 
from \citet[][purple squares]{esteban09,esteban20} and \citet[][blue pentagons]{skillman20}. 
The dashed line is the least-squares fit from \citet{garnett95},
while the  blue curve shows the empirical stellar curve from \citet{nicholls17}.
Some of our preliminary CEL C/O measurements (turquoise diamonds) agree with the overall 
increasing C/O--O/H trend, while others appear to be low-C/O outliers, but are 
not significant enough to discern from the large scatter.
\label{fig12}}
\end{center}
\end{figure}


\subsubsection{C/O Comparison to Previous Works}\label{comparison}
Several previous studies have analyzed the nebular abundances in M101. 
Of these studies, four are particularly relevant to this work: 
(1) \citet{garnett99} observed three \ion{H}{2} regions in M101 using the Faint 
Object Spectrograph (FOS) on {\it HST}. 
\ion{C}{3}] \W\W1907,1909 was detected significantly in all three, 
but only upper limits could be placed on \ion{O}{3}] \W\W1661,1666. 
As tabulated in Table~\ref{tbl7}, two of these regions overlap with our 
\ion{H}{2} region sample in M101 (NGC5455, NGC5471). 
(2) \citet{esteban09} obtained high-resolution optical spectroscopy with the 
High Resolution Echelle Spectrometer \citep[HIRES;][]{vogt94} on the Keck I telescope 
for two \ion{H}{2} regions in M101 and intermediate-resolution spectroscopy 
of NGC 5447 using the Intermediate dispersion Spectrograph and Imaging System 
(ISIS) spectrograph on the 4.2m WHT.
They measured significant \ion{C}{2} RLs in H1013 and NGC 5447b, 
which overlaps with our UV CEL measurement.
(3) \citet{skillman20} measured significant \ion{C}{2} RLs in 10 \ion{H}{2} regions 
in M101 using low-resolution optical spectroscopy taken with the MODS on the LBT.
(4) \citet{esteban20} used low-resolution optical spectroscopy taken with the 
Optical System for Imaging and low-Intermediate-Resolution Integrated Spectroscopy 
\citep[OSIRIS;][]{cepa00} spectrograph on the Gran Telescopio Canarias (GTC) to measure 
the RLs in 11 \ion{H}{2} regions in M101, five of which have significant \ion{C}{2}.
Of these regions, NGC 5455 and NGC 5471 are in common with \citet{garnett99}, 
\citet{skillman20}, and this work, NGC 5447b is in common with \citet{esteban09}, 
\citet{skillman20}, and this work, H1013 overlaps with the \citet{skillman20} sample, 
while H37 has no comparison. 
These previous works provide consistency checks on our measurements.

We conduct a preliminary comparison of our UV C/O abundance measurements to previous
studies in the C/O versus O/H plot in Figure~\ref{fig12}.
In general, the study of the C/O versus O/H trend is of interest to improve our 
understanding of C production and the conditions in which C emission is strongest
(e.g., high-redshift star-forming regions). 
Theoretically, we expect C/O nucleosynthetic production to be constant via the 
triple-$\alpha$ process. 
However, observationally an increasing trend in C/O with O/H is clearly visible,
but with significant scatter.
This dispersion has prevented us from further discerning whether C/O increases
linearly or gradually with O/H (e.g., the dashed line from \cite{garnett95}
or solid blue line from \cite{nicholls17} in Figure~\ref{fig12}) or has a 
bimodal trend with a plateau at low O/H, similar to the N/O trend. 

A further challenge to understanding the C/O--O/H relationship is the differing sources
of C/O measurements; at low-metallicity C/O is measured via the collisionally-excited 
UV lines, while C/O is typically measured via the optical RLs at high-metallicity.
Therefore, our measurements, combined with the CEL measurements from \citet{garnett99},
provide an important extension of the CEL measurements at low-metallicity that would
allow an apples-to-apples assessment of the C/O--O/H trend if more significant 
measurements were obtained.
Along these lines, our measurements agree well with the dispersion of the C/O trend in 
Figure~\ref{fig12}, but also lie at low C/O ratios on average.
In particular, the low-metallicity (gold diamonds) plateau has an average 
log(C/O)$_{\rm CEL} = -0.71$, while the high-metallicity average 
(\cite{garnett99} plus this work) log(C/O)$_{\rm CEL} = -0.54$.
Note that this value could be largely biased by the low average of our seven points 
(log(C/O)$_{\rm CEL, B24} = -0.70$), whereas the \citet{garnett99} sample alone has 
a much higher average (log(C/O)$_{\rm CEL, G99} = -0.36$).
In comparison, the high-metallicity average log(C/O)$_{\rm RL} = -0.29$, or roughly
0.1 dex higher than previous CEL C/O measurements, and much higher than the
values estimated here.
Despite the large uncertainties that remain, these results suggest that 
(1) the increasing C/O trend is real and not just an artifact
of different measurement methods and 
(2) that a C/O abundance discrepancy factor (ADF) 
exists on the same order of magnitude, or larger, as the O/H ADF.
However, little more can be said given the large uncertainties on our C/O measurements.
Further progress to improve comparison of C/O from the UV CELs to those from optical RLs 
and better constrain the source(s) of the ADF will require higher S/N and higher 
spectral-resolution spectra.


\begin{deluxetable*}{rlcCCcc}
\tabletypesize{\scriptsize}
\tablecaption{Comparison of M101 CEL and RL C/O Measurements}
\tablehead{
\multicolumn{2}{c}{Name} & \CH{R.A., Decl.} & \CH{12+log(O/H)} & \CH{log(C/O)}  & \CH{Method} & \CH{Source}}
\startdata	
H1013$^1$  & Reg 1 & 14:03:31.3, +54:21:05.8 & 8.57\pm0.02 & -0.04\pm0.07 & RL  & \citet{skillman20} \\
           &       & 14:03:31.2, +54:21:14.8 &             & +0.22\pm0.13 & RL  & \citet{esteban09}  \\ \hline
H1052$^1$  & Reg 2 & 14:03:34.1, +54:18:39.6 & 8.57\pm0.01 & -0.08\pm0.04 & RL  & \citet{skillman20} \\ \hline
NGC~5461   & Reg 3 & 14:03:41.6, +54:19:08.5 & 8.48\pm0.02 & -0.31\pm0.12 & RL  & \citet{skillman20} \\ 
           &       & 14:03:41.6, +54:19:04.4 &             & -0.03\pm0.13 & CEL & \citet{garnett99}$^2$  \\ \hline
NGC~5462a  & Reg 4 & 14:03:53.0, +54:22:06.8 & 8.45\pm0.05 & -0.39\pm0.19 & RL  & \citet{skillman20} \\
           &       &                         &             & {\it -0.79}  & CEL & This work          \\ \hline
NGC~5462b  & Reg 5 & 14:03:53.8, +54:22:10.8 & 8.43\pm0.01 & -0.28\pm0.12 & RL  & \citet{skillman20} \\ 
           &       &                         &             & {\it -0.51}  & CEL & This work          \\ \hline
           &       & 14:03:53.1, +54:22:06.4 &             & -0.12\pm0.10 & RL  & \citet{esteban20}  \\ \hline
NGC~5447   & Reg 6 & 14:02:30.5, +54:16:09.9 & 8.42\pm0.01 & -0.30\pm0.09 & RL  & \citet{skillman20} \\ 
           &       &                         &             & {\it -0.81}  & CEL & This work          \\ \hline
NGC~5455   & Reg 7 & 14:03:01.1, +54:14:28.0 & 8.39\pm0.02 & -0.34\pm0.11 & RL  & \citet{skillman20} \\
           &       &                         &             & {\it -0.45}  & CEL & This work          \\
           &       & 14:03:01.2, +54:14:27.0 &             & -0.26\pm0.14 & CEL & \citet{garnett99}$^2$  \\
           &       & 14:03:01.2, +54:14:29.4 &             & -0.22\pm0.11 & RL  & \citet{esteban20}  \\ \hline
NGC~5447   & Reg 8 & 14:02:27.8, +54:16:25.3 & 8.35\pm0.01 & -0.25\pm0.17 & RL  & \citet{skillman20} \\
           &       &                         &             & {\it -0.87}  & CEL & This work          \\ 
           &       & 14:02:28.7, +54:16:25   &             & -0.15\pm0.12 & RL  & \citet{esteban09}  \\
           &       & 14:02:28.1, +54:16:26.9 &             & -0.27\pm0.11 & RL  & \citet{esteban20}  \\ \hline
H37$^1$    &       & 14:02:17.7, +54:22:34.1 & 8.30\pm0.01 & -0.14\pm0.14 & RL  & \citet{esteban20}  \\ \hline
H1216$^1$  & Reg 9 & 14:04:10.9, +54:25:19.2 & 8.26\pm0.03 & -0.63\pm0.50 & RL  & \citet{skillman20} \\
           &       &                         &             & {\it -0.85}  & CEL & This work          \\ \hline
NGC~5471   & Reg 10& 14:04:29.0, +54:23:48.6 & 8.14\pm0.03 & -0.47\pm0.46 & RL  & \citet{skillman20} \\
           &       &                         &             & -0.65\pm0.16 & CEL & This work          \\
           &       & 14:04:29.0, +54:23:48.8 &             & -0.67\pm0.05 & CEL & \citet{garnett99}$^2$  \\
           &       & 14:04:29.0, +54:23:49.0 &             & -0.20\pm0.11 & RL  & \citet{esteban20}  
\enddata	
\tablecomments{
Comparison of M101 CEL C/O abundances to RL C/O abundances from the literature and this work.
Column 1 lists the common name of the \ion{H}{2} region in M101, along with the 
region name of this work, while coordinate R.A. and Decl. (J2000) is given in Column 2.
Column 3 lists the measured 12$+$O/H abundances from \citet{croxall16}, 
while Column 4 lists the C/O abundance measured using the method in Column 5 by the reference in Column 6.
Note that estimates of log(C/O) using insignificant detections on \ion{O}{3}] in this work are
identified with italicized values. \\
$^1$Name adopted from the catalog of \citet{hodge90}.
$^2$Note that the values from \citet{garnett99} are those calculated with an R$_V$ value of 3.1. 
\citet{garnett99} also calculated values assuming an R$_V$ value of 5, which can give significantly 
different results.}
\label{tbl7}
\end{deluxetable*}


\section{Conclusions}\label{sec:conclusions}
We present low-resolution {\it HST}/COS FUV spectroscopy of 9 \ion{H}{2} regions in 
the star-forming disk of spiral galaxy M101. 
These regions were selected for their high surface brightnesses and existing direct 
abundances and \ion{C}{2} \W4267 RL detections from the optical CHAOS survey,
with the goal of detecting the FUV \ion{O}{3}] \W1666 and \ion{C}{3}] 
\W\W1907,1909 CELs.
Strong \ion{C}{3}] emission lines were detected for the 7 lowest-metallicity \ion{H}{2} 
regions in our sample (12+log(O/H)$\lesssim8.5$), but \ion{O}{3}] \W1666 was significantly 
detected in only the most metal-poor region (Reg 10; 12+log(O/H) = 8.16).
The other six \ion{H}{2} regions have tentative \ion{O}{3}] detections that would be
improved with higher spectral resolution due to the narrow intrinsic widths of the lines
and the contamination of the Milky Way \ion{Al}{2} \W1671 ISM absorption line. 

We performed an exploratory C/O abundance analysis with the limited \ion{O}{3}] and 
\ion{C}{3}] line detections.
To convert the C/O emission line ratios to relative abundances, we produced new analytic 
functions of the carbon ICF as a function of metallicity that are calibrated to 
the conditions of the M101 \ion{H}{2} regions. 
To do so, we created a custom grid of \texttt{CLOUDY} photoionization models 
using \texttt{BPASS} single-burst models as inputs.
With our tentative C/O abundance measurements, we compared to the existing sample of C/O 
measurements for M101, consisting of 17 optical RL measurements and 3 FUV CEL measurements for 
11 different \ion{H}{2} regions (see Table~\ref{tbl7}).
Our preliminary C/O measurements add 7 CEL points to this sample, but five of these
points our low C/O outliers from the general C/O--O/H trend (Figure~\ref{fig12}).
While the FUV C/O abundances can provide useful constraints on the chemical
enrichment of galaxies, higher-S/N and higher-resolution CEL measurements are clearly 
needed to robustly constrain the high-metallicity end of the C/O--O/H and 
compare to RL C/O abundances.

Despite the lack of strong \ion{O}{3}] CELs, the resulting FUV spectra revealed 
numerous continuum features, such as the massive-star wind features, that are useful 
for characterizing the ionizing stellar population.
We fit stellar continuum with both \texttt{SB99} and \texttt{BPASS} 
luminosity-weighted stellar burst models to determine the characteristic 
age and metallicity of the massive star populations, finding
a range of ages (2--10 Myr) and metallicities (0.5--1.0 $Z_\odot$) that 
seem to correlate with the He and N emission and P-Cygni profiles observed.
In particular, the youngest stellar populations are strongly correlated with 
broad Wolf-Rayet (WR) \ion{He}{2} \W1640 emission and enhanced N/O abundances
(log(N/O)$>-1.2$), as determined from the optical 
[\ion{N}{2}] \W6584/[\ion{O}{2}] \W3727 ratio. 
Detections of WR stars are further supported by visual confirmation of the 
WR ``blue bump" feature around \W4650 in the optical spectra, where the strong
\ion{N}{2} \W\W4634,4641,4643 emission is indicative of nitrogen-type WR (WN) stars.
As a result, this special phase of stellar evolution may be responsible for the
\ion{N}{4} \W\W1483,1486 emission and the excess emission in the \ion{N}{4} \W1718 
stellar wind feature, which in turn biases the \texttt{SB99} and \texttt{BPASS} 
stellar continuum fits to higher metallicities relative to the gas-phase metallicities. 
Therefore, the WR \ion{H}{2} regions of M101 presented here provide a strong test-bed 
to constrain future WR atmosphere and evolution models, and their incorporation to 
stellar population models.

Finally, we observed additional P-Cygni and emission features in the $\sim900-1200$ \AA\
regime of our M101 spectra that further support strong contributions from WN stars
(e.g., \ion{S}{6} \W\W933,945, \ion{N}{4} \W955, \ion{N}{3} \W991, \ion{O}{6} \W\W1032,1038, 
\ion{P}{5} \W\W1118,1128, and \ion{C}{3} \W1175), and note a lack of strong
\ion{C}{3} and \ion{C}{4} features in this regime (e.g., \ion{C}{3} \W977, 
\ion{C}{4} \W1197, \ion{C}{4} \W1135, \ion{C}{3} \W1140, \ion{C}{4} \W1169) 
that would be expected if WC stars were present.
Therefore, we recommend the $\sim 900-1200$ \AA\ stellar continuum as a powerful 
diagnostic regime for identifying the presence of WR stars in the integrated spectra 
of star-forming regions and galaxies.

\facilities{HST (COS), LBT (MODS)}
\software{
\texttt{astropy} \citep{astropy:2013, astropy:2018, astropy:2022},
\texttt{CalCOS} (STScI),
\texttt{dustmaps} \citep{green18},
\texttt{jupyter} \citep{kluyver16},
\texttt{LINMIX} \citep{kelly07}, 
\texttt{MPFIT} \citep{markwardt09},
MODS reduction Pipeline,
\texttt{Photutils} \citep{bradley21},
\texttt{LMFIT} version 1.2.2 \citep{newville18},
\texttt{PyNeb} version 1.1.14 \citep{luridiana15},
\texttt{BPASS} version 2.2 \citep{stanway18},
\texttt{Starburst99} \citep{leitherer14},
\texttt{numpy} version 1.26 \citep{harris20}}

\begin{acknowledgements}
The authors are grateful to the referee for insightful comments and questions that 
led to better parameter characterization and an overall improved paper.
DAB is grateful for the support for this program, HST-GO-15126, that was provided by 
NASA through a grant from the Space Telescope Science Institute, 
which is operated by the Associations of Universities for Research in Astronomy, 
Incorporated, under NASA contract NAS5-26555. 

This work was based in part on observations made with the Large Binocular Telescope (LBT). The LBT is an international collaboration among institutions in the United States, Italy, and Germany. The LBT Corporation partners are: the University of Arizona on behalf of the Arizona university system; the Istituto Nazionale di Astrofisica, Italy; the LBT Beteiligungsgesellschaft, Germany, representing the
Max Planck Society, the Astrophysical Institute Potsdam, and
Heidelberg University; the Ohio State University; and the
Research Corporation, on behalf of the University of Notre
Dame, the University of Minnesota, and the University of
Virginia. This research used the facilities of the Italian Center for Astronomical Archive (IA2) operated by INAF at the Astronomical Observatory of Trieste.

This paper made use of the modsIDL spectral data reduction reduction pipeline
developed in part with funds provided by NSF Grant AST-1108693 and a generous
gift from OSU Astronomy alumnus David G. Price through the Price Fellowship in
Astronomical Instrumentation. 

All the {\it HST} data used in this paper can be found in MAST: 
\dataset[ 10.17909/qv80-ph56]{http://dx.doi.org/ 10.17909/qv80-ph56}.
\end{acknowledgements}


\bibliographystyle{aasjournal}
\bibliography{mybib}

\clearpage

\end{document}